\documentclass[lettersize,journal]{IEEEtran}
\usepackage{amsmath,amsfonts}
\usepackage{array}
\usepackage[caption=false,font=normalsize,labelfont=sf,textfont=sf]{subfig}
\usepackage{algpseudocode}
\usepackage{graphicx}
\usepackage{caption}
\usepackage{booktabs}
\usepackage{textcomp}
\usepackage{stfloats}
\usepackage{url}
\usepackage{cite}
\usepackage{verbatim}
\usepackage[ruled,linesnumbered]{algorithm2e}
\begin{document}

\title{Towards Scalable Quantum Key Distribution: A Machine Learning-Based Cascade Protocol Approach}

\author{Hasan Abbas Al-Mohammed, Saif Al-Kuwari, Hashir Kuniyil, Ahmed Farouk
\thanks{H. A. Al-Mohammed is with the Department of Computer Science and Engineering, Qatar University, Doha, Qatar. E-mail: ha1800217@qu.edu.qa}
\thanks{S. Al-Kuwari and Hashir Kuniyil are with the Qatar Center for Quantum Computing, College of Science and Engineering, Hamad Bin Khalifa University, Doha, Qatar. E-mail: smalkuwari@hbku.edu.qa, hkuniyil@hbku.edu.qa}
\thanks{A. Farouk is with the Department of Computer Science, Faculty of
Computers and Artificial Intelligence, South Valley University, Hurghada,
Egypt. e-mail: (ahmed.farouk@sci.svu.edu.eg).}
\thanks{This work was partially supported by the Qatar National Research Fund (a member of the Qatar Foundation) under NPRP Grant number 12S-0228-190177.}
}

\maketitle

%\markboth{IEEE Transactions on Wireless Communications,~Vol.~XX, No.~X, XXXX~XXXX}
%{Al-Mohammed \MakeLowercase{\textit{et al.}}: Towards Scalable Quantum Key Distribution: A Machine Learning-Based Cascade Protocol Approach}

\begin{abstract}

Quantum Key Distribution (QKD) is a pivotal technology in the quest for secure communication, harnessing the power of quantum mechanics to ensure robust data protection. However, scaling QKD to meet the demands of high-speed, real-world applications remains a significant challenge. Traditional key rate determination methods, dependent on complex mathematical models, often fall short in efficiency and scalability. In this paper, we propose an approach that involves integrating machine learning (ML) techniques with the Cascade error correction protocol to enhance the scalability and efficiency of QKD systems.  Our ML-based approach utilizes an autoencoder framework to predict the Quantum Bit Error Rate (QBER) and final key length with over 99\% accuracy. This method significantly reduces error correction time, maintaining a consistently low computation time even with large input sizes, such as data rates up to 156 Mbps. In contrast, traditional methods exhibit exponentially increasing computation times as input sizes grow, highlighting the superior scalability of our ML-based solution.  Through comprehensive simulations, we demonstrate that our method not only accelerates the error correction process but also optimizes resource utilization, making it more cost-effective and practical for real-world deployment. The Cascade protocol's integration further enhances system security by dynamically adjusting error correction based on real-time QBER observations, providing robust protection against potential eavesdropping. 
 Our research establishes a new benchmark for scalable, high-throughput QKD systems, proving that machine learning can significantly advance the field of quantum cryptography. This work continues the evolution towards truly scalable quantum communication.
\end{abstract}


\section{Introduction}
\IEEEPARstart{A}{s} we transition into the era of 6G and beyond, the demand for secure and efficient data transmission has become increasingly critical \cite{nguyen2021security}. Quantum Key Distribution (QKD) presents a groundbreaking solution for ultra-secure communication by harnessing the unique features of quantum mechanics \cite{al2021use, jiang2021road}. However, for QKD to become practical and be widely adopted, several challenges must be overcome.

one of the major challenges being addressed in the literature is achieving a high data rate. Traditional QKD systems often suffer from low transmission rates, which restrict their use in high-speed communication networks. By implementing high data rate protocols and employing advanced hardware, we can significantly enhance transmission speeds \cite{ecker2021strategies}. Modern optical communication technologies, when integrated with QKD systems, can achieve data rates in the hundreds of megabits per second, making QKD a viable option for real-world applications \cite{chapuran2009optical, wang2017long}.

Error correction efficiency is another critical factor \cite{liu2017experimental}. Quantum channels are prone to errors due to noise and other environmental factors, which can compromise the reliability of key distribution. Advanced error correction protocols, such as the Cascade protocol, can address this issue effectively \cite{tupkary2023using}. By incorporating machine learning (ML) techniques, we can further improve the robustness and efficiency of error correction. ML models can predict the Quantum Bit Error Rate (QBER) and final key length with high accuracy, allowing for more effective error correction with fewer iterations and faster convergence.

Scalability is a major hurdle for QKD networks. The complexity and resource requirements also rise as the number of users and the distances between them increase. Developing scalable network architectures, such as quantum repeaters\cite{briegel1998quantum} and satellite-based QKD \cite{dai2020towards}, can extend the range and capacity of QKD systems. Additionally, ML algorithms can optimize network resources and routing, enhancing the scalability of QKD networks \cite{fiandrino2020machine, zhao2018resource}.

Ensuring the security of QKD against various types of attacks is paramount. This includes attacks targeting both the physical layer and the post-processing steps \cite{cao2022evolution}. Continuous advancements in quantum cryptography protocols, combined with rigorous security analysis and testing, can strengthen the security of QKD systems. Implementing ML techniques for real-time threat detection and response can further enhance the security of quantum communication networks \cite{al2021machine}.

Cost and resource optimization are also crucial for the widespread deployment of QKD. The high cost and resource requirements of quantum communication equipment can be a barrier. Advances in quantum technology, such as the development of cost-effective quantum sources and detectors, can help reduce overall costs \cite{alleaume2009topological}. Furthermore, using ML for resource optimization can ensure efficient allocation and utilization of quantum communication resources \cite{wallnofer2020machine}.

By addressing these challenges through targeted advancements, we can achieve practical communication for QKD, paving the way for secure and efficient quantum communication networks. These improvements will enhance QKD's feasibility and ensure its robustness, scalability, and seamless integration with existing communication systems.

In this paper, we propose a novel approach to enhance the efficiency and practicality of QKD by integrating machine learning (ML) techniques with traditional QKD protocols. Specifically, we address the challenge of error correction by employing a machine learning model to predict the Quantum Bit Error Rate (QBER) and the final key length, thereby optimizing the error correction process.

Traditional methods for error correction in QKD, such as the Cascade protocol, involve iterative processes that can be computationally intensive and time-consuming \cite{tupkary2023using,mao2022high}. These methods require extensive calibration and tuning to account for various noise levels and environmental factors, leading to significant delays, especially when dealing with large datasets. While these traditional methods are effective, they often result in high computational costs and prolonged convergence times \cite{gumucs2021novel,mao2022high}.
%start
Our approach leverages the power of machine learning to streamline and accelerate this process. By training an ML model on historical data, we can predict the QBER and final key length with high accuracy. This prediction allows us to implement error correction more efficiently, reducing the number of iterations needed and achieving faster convergence. The use of ML also enhances the adaptability of the system, enabling it to handle varying noise levels and environmental conditions without extensive recalibration.

The primary contributions of our work include:

\begin{itemize}
    \item \textbf{Improved Error Correction Efficiency}: By integrating ML with the Cascade protocol, we significantly enhance the robustness and speed of error correction. Our ML model accurately predicts QBER, leading to fewer iterations and quicker convergence compared to traditional methods.
    
    \item \textbf{Scalability and High Data Rates}: Our method is designed to handle high data rates, such as 156 Mbps, and can scale efficiently with the size of the dataset. This makes our approach suitable for modern high-speed communication networks, where traditional methods might struggle with scalability and speed.
    
    \item \textbf{Cost and Resource Optimization}: The use of ML allows for better resource allocation and utilization, reducing the overall computational cost. This makes our method more cost-effective and practical for real-world implementations.
    
    \item \textbf{Enhanced Security}: The integration of ML for real-time prediction and error correction also improves the security of the QKD system. By quickly adapting to changing conditions and potential threats, our method ensures a higher level of security for quantum communication.
\end{itemize}

In comparison to traditional methods, our ML-enhanced approach demonstrates superior performance in terms of speed, efficiency, and adaptability. While traditional QKD protocols rely on extensive computations and iterations, our method leverages predictive capabilities to streamline the process, making QKD more practical and feasible for widespread adoption in secure communication networks.

Therfore, our proposed method not only addresses the inherent challenges of traditional QKD systems but also paves the way for more efficient, scalable, and secure quantum communication. By harnessing the power of machine learning, we bring QKD one step closer to practical, real-world applications.
%end   

\section{Related Work}

The field of QKD has undergone significant advancements since the inception of foundational protocols such as BB84 \cite{bennett2014quantum} and E91 \cite{ekert1991quantum}. These traditional QKD protocols leverage the principles of quantum mechanics to ensure secure key exchange. However, they encounter limitations in terms of error correction, scalability, and data rates \cite{hong2016challenges}. The BB84 protocol, introduced by Bennett and Brassard, and the E91 protocol, proposed by Ekert, are among the most widely recognized QKD protocols.BB84 is a prepare and measure protocol which employs single photons in four polarization states, whereas E91 relies on quantum entanglement. Despite their theoretical security, these protocols necessitate robust error correction mechanisms to counteract the noise and errors inherent in quantum channels \cite{xu2020secure}.

The Cascade protocol is a commonly used error-correction method in QKD \cite{brassard1993secret}. It iteratively identifies and rectifies errors through a series of interactive steps between the communicating parties. While effective, Cascade and other traditional methods such as LDPC (Low-Density Parity-Check) codes are computationally demanding and time-consuming, especially when dealing with large datasets and high data rates \cite{xu2020secure}.

Recent advancements have explored the application of machine learning (ML) techniques to enhance various aspects of quantum communication systems \cite{ding2020predicting, lu2019parameter}. ML models have been utilized to predict noise for long-distance communication \cite{huang2016long}, optimize key rate estimation \cite{wang2019machine}, and improve error correction processes \cite{li2018discrete, wallnofer2020machine}. Studies have demonstrated that ML can adapt to changing environmental conditions and provide more accurate predictions compared to traditional methods.

Several attempts have been made to integrate ML with QKD protocols. For instance, ML techniques have been employed to predict the Quantum Bit Error Rate (QBER) and optimize the error correction process \cite{wang2019machine, dunjko2018machine}. Our work builds on these efforts by leveraging an autoencoder-based approach to enhance the Cascade protocol, thereby achieving faster and more efficient error correction.

Comparative analyses have shown that ML-enhanced QKD systems can outperform traditional methods in terms of computational efficiency and scalability \cite{liu2022automated}. Studies comparing the performance of ML-based error correction with traditional methods have reported significant reductions in convergence time and computational complexity \cite{li2018discrete}. These advancements underscore the potential of ML to revolutionize QKD by addressing its longstanding challenges, paving the way for more practical and scalable quantum communication systems.

Recent advancements in Quantum Key Distribution (QKD) have seen significant contributions from the integration of machine learning techniques to enhance security and efficiency. One notable study \cite{al2021machine} addresses the challenge of detecting attackers in IoT networks during QKD without disrupting the key distribution process by employing artificial neural networks and deep learning, achieving 99\% accuracy in practical railway scenarios. Another innovative approach \cite{mao2020detecting} proposes a defense strategy for Continuous-Variable QKD (CVQKD) systems using an artificial neural network model to detect and classify various attack types, establishing a universal attack detection method. Additionally, research by \cite{liu2019practical} explores using Long Short-Term Memory (LSTM) networks to predict phase modulation variations, maintaining low quantum-bit error rates and enhancing key transmission efficiency through real-time control. \cite{liu2018integrating} present a support vector regression model to predict time-evolving physical parameters in CVQKD systems, optimizing performance and security without the need for additional resources. Furthermore, \cite{chin2021machine} demonstrate a machine learning framework based on the Unscented Kalman Filter (UKF) for accurate phase noise estimation and compensation in CVQKD systems, ensuring stable performance with reduced hardware complexity. Collectively, these studies highlight the potential of machine learning to advance the practical implementation and security of QKD systems significantly.
%end

\section{BACKGROUND AND METHODOLOGY}

In this section, we describe the asymptotic formulation of the BB84 protocol using the Cascade error correction protocol during the information reconciliation step, and detail the methodology of integrating the ML to predict QBER and optimize error correction.

\subsection{List of Variables and Their Definitions}
Table ~\ref{tab:variables} contains all the variables used in this paper with their definitions, to make the paper clear and consistent.

\begin{table}[]
\centering
\caption{List of Variables and Their Definitions}
\label{tab:variables}
\begin{tabular}{|c|p{6cm}|}
\hline
\textbf{Variable} & \textbf{Definition} \\
\hline
$QKD$ & Quantum Key Distribution \\
\hline
$BB84$ & A QKD protocol introduced by Bennett and Brassard \\
\hline
$E91$ & A QKD protocol proposed by Ekert, based on quantum entanglement \\
\hline
$QBER$ & Quantum Bit Error Rate; the ratio of erroneous bits to total bits transmitted \\
\hline
$\delta^{A}_{\text{leak}}$ & Analytical upper bound on the number of bits sent from Alice to Bob per bit of raw key \\
\hline
$h(e)$ & Binary entropy function \\
\hline
$f$ & Efficiency factor, typically ranging between 1 and 1.5 \\
\hline
$R_{\text{incorrect}}$ & Incorrectly calculated key rate due to not accounting for Bob to Alice communication \\
\hline
$S(Z \mid E \tilde{A} \tilde{B})$ & Conditional von Neumann entropy of $Z$ given Eve's information $E$ and classical announcements $\tilde{A} \tilde{B}$ \\
\hline
$p_{\text{pass}}$ & Probability that the signal passes sifting \\
\hline
$\delta_{\text{leak}}$ & Number of bits used during error correction per bit of raw key \\
\hline
$R$ & Key rate \\
\hline
$T_{\text{traditional}}$ & Overall computational complexity of traditional methods \\
\hline
$T_{\text{autoencoder}}$ & Overall computational complexity of the autoencoder approach \\
\hline
$n$ & Input size (number of bits or data size) \\
\hline
$h$ & Size of the hidden layer in the autoencoder \\
\hline
$K_i$ & Kraus operators representing measurements, announcements, and sifting by Alice and Bob \\
\hline
$Z_j$ & Kraus operators implementing a pinching channel on the key register \\
\hline
$D(X \parallel Y)$ & Quantum relative entropy between $X$ and $Y$ \\
\hline
$G(\rho)$ & Intermediate state in the optimization problem \\
\hline
$F$ & Objective function for optimization \\
\hline
$\alpha, \beta$ & Alice and Bob's declarations (like basis selection) \\
\hline
$P_A(\alpha, x), P_B(\beta, y)$ & Positive Operator-Valued Measures (POVMs) representing measurement outcomes by Alice and Bob \\
\hline
$\tilde{A}, \tilde{B}$ & Classical declarations made by Alice and Bob \\
\hline
$\rho$ & Quantum state \\
\hline
$\gamma$ & Expectation values obtained during parameter estimation \\
\hline
$r(\alpha, \beta, x)$ & Key map implemented by Alice \\
\hline
$\mathcal{E}_{\text{misalign}}(\rho)$ & Misalignment operation on quantum state $\rho$ \\
\hline
$\mathcal{E}_{\text{depol}}(\rho)$ & Depolarization operation on quantum state $\rho$ \\
\hline
$R_r$ & Repetition rate \\
\hline
$f(\rho)$ & Function used in the optimization problem for $F$ \\
\hline
$x, y$ & Bits representing measurement outcomes by Alice and Bob \\
\hline
$w$ & Additional announcement representing $x \oplus y$ \\
\hline
$\eta$ & Learning rate for the Adam optimizer \\
\hline
$I$ & Mini-batch indices for training \\
\hline
$\hat{y}$ & Predicted output of the autoencoder \\
\hline
$L$ & Loss function (mean squared error) \\
\hline
$\theta$ & Parameters of the autoencoder model \\
\hline
$W, b$ & Weight matrix and bias vector in the autoencoder \\
\hline
$\lambda$ & Regularization parameter \\
\hline
$L1$ & L1-norm regularization term \\
\hline
$f(x)$ & Linear decoding transfer function used in the autoencoder \\
\hline
$y_i, \hat{y}_i$ & True and predicted values in the mean squared error calculation \\
\hline
$\text{decoded\_output}$ & Output of the autoencoder after decoding \\
\hline
$\text{corrected\_output}$ & Output after applying the Cascade protocol for error correction \\
\hline
$\text{final\_output}$ & Final output after applying the trained autoencoder to the corrected data \\
\hline
$a_n, b_n$ & Fourier series coefficients for representing the relationship between sender device and final key length \\
\hline
$\textbf{\( F \)}$ & The objective function to be minimized, defined as the minimum quantum relative entropy between the quantum operation \( G(\rho) \) and its projection \( Z(G(\rho)) \) over all density matrices \( \rho \) in the state space \( \mathcal{S}(\gamma) \).
\\
\hline
$\textbf{\( F' \)}$ & The objective function in the optimization problem that minimizes the quantum relative entropy between the quantum operations \( G(\rho) \) and \( Z(G(\rho)) \) applied to the state \( \rho \).

\\
\hline
\end{tabular}
\end{table}

\subsection{Protocol description }
To integrate the Cascade protocol for QKD and implement our ML-based approach, we utilized the enhanced Cascade protocol. This enhancement addresses a key weakness in evaluating QKD protocols using the bidirectional error-correction Cascade. Building on the work in \cite{tupkary2023using}, which focused on security, our approach not only enhances the latest version of the Cascade protocol but also aims to predict the final key length, distinguishing it from previous research.

\subsubsection{Quantum Phase}

In the BB84 protocol, Alice prepares qubits in one of four polarization states: 0°, 90°, 45°, and 135° (or -45°). These states are grouped into two basis sets:

\begin{itemize}
    \item \textbf{Rectilinear Basis (\(|+\rangle\))}: Consists of 0° (representing bit 0) and 90° (representing bit 1).
    \item \textbf{Diagonal Basis (\(|\times\rangle\))}: Consists of 45° (representing bit 0) and 135° (or -45°) (representing bit 1).
\end{itemize}

The states can be mathematically represented as:

\[
|0\rangle, \quad |1\rangle, \quad |+\rangle = \frac{1}{\sqrt{2}}(|0\rangle + |1\rangle), \quad |-\rangle = \frac{1}{\sqrt{2}}(|0\rangle - |1\rangle)
\]

In this scheme, Alice sends a random bit in a random basis, and Bob measures in a random basis. The four states are treated as two bases, each with two possible outcomes (0 and 1).

According to the source-replacement method \cite{curty2004entanglement}, one can describe the protocol to be equivalent as follows: Alice produces the Bell state \(\left|\psi\right\rangle_{AA'} = \left|\phi_+\right\rangle = \frac{|00\rangle + |11\rangle}{\sqrt{2}}\) and transmits \(A'\) to Bob. to model the misalignment we can write  as a rotation by an angle \(\theta\) around the Y axis on \(A'\), with:

\begin{equation}
U(\theta) = I_A \otimes \begin{pmatrix}
\cos(\theta) & -\sin(\theta) \\
\sin(\theta) & \cos(\theta)
\end{pmatrix}
\end{equation}
The effect of misalignment on the quantum state \(\rho\) is given by:
\begin{equation}
\mathcal{E}_{\text{misalign}}(\rho) = U(\theta) \rho U(\theta)^{\dagger}
\end{equation}

The depolarization process is represented as follows, where \( q \) denotes the probability that the quantum state undergoes depolarization:

\begin{equation}
\mathcal{E}_{\text{depol}}(\rho) = (1 - q)\rho + q \, \text{Tr}_{A'}(\rho) \otimes \frac{I_{B}}{2}
\end{equation}

The state on which statistical methods are calculated is:

\begin{equation}
\rho_{AB} = \mathcal{E}_{\text{depol}}(\mathcal{E}_{\text{misalign}}(|\phi_+\rangle \langle \phi_+|))
\end{equation}

In this context, \( q \) controls the extent to which the quantum state \( \rho \) is affected by the depolarization process.

Both Bob as well as Alice make measurements on qubit systems, and their Positive Operator-Valued Measures (POVMs) are:
\begin{align}
P_{Z,0} &= p_{z} |0\rangle \langle 0|, P_{Z,1} = p_{z} |1\rangle \langle 1|, \nonumber \\
P_{X,0} &= p_{x} |+\rangle \langle +|, P_{X,1} = p_{x} |-\rangle \langle -|
\end{align}
using \(p_{z} = p_{x} = 1/2\). Alice creates the keymap by simply placing the measurement results into the corresponding key register.

Using the source-replacement scheme \cite{curty2004entanglement}, specific registers formed by the generic version of the Kraus operators are capable of being eliminated to simplify the code. Specifically, the improved technique takes into account the registers that hold Alice and Bob's measurement results, necessitating only one duplicate of the announcement register. This simplification gives the overall structure used by Kraus operators as follows:

\begin{equation}
K_{\alpha} = \sum_{x} |r(\alpha, x)\rangle_Z \otimes \sum_{y} P^{A}_{x\alpha}(y) \otimes P^{\beta}_{\alpha}(y) \otimes |x\rangle_A,
\end{equation}
which becomes:
\begin{align}
K_{\alpha, w} &= \sum_{x} |r(\alpha, x)\rangle_Z \otimes \sum_{y \mid x \oplus y = w} P^{A}_{x\alpha}(y) \otimes P^{\beta}_{\alpha}(y) \nonumber \\
&\quad \otimes |x\rangle_A \otimes |w\rangle_W.
\end{align}
Here, \(\alpha\) and \(\beta\) denote the basis choices, and \(x\) and \(y\) indicate the measurement results.
The variable \(w\) represents the error syndrome, calculated as the parity information \(w = x \oplus y\). Alice and Bob’s POVMs are indicated by \(P^{A} = P^{A}_{x\alpha}\) and \(P^{\beta} = P^{\beta}_{\alpha y}\). When Alice and Bob eliminate all transmissions with basis mismatches, the set of operators that produce the \(G\) map is \(\{K_{\alpha}\}\), and the group of procedures that generate \(G'\) is \(\{K_{\alpha, w}\}\). The \(Z\) map has Kraus operators \(\{Z_j\}\) given by \(Z_j = |j\rangle \langle j |_Z \otimes I_{ABX\tilde{A}\tilde{B}}\). Thus, the last Kraus operators of \(F\):
\begin{equation}
\begin{aligned}
K_{Z} &= \left( \begin{pmatrix}
0 \\
\sqrt{\frac{1}{2}}
\end{pmatrix} \otimes \sqrt{\frac{1}{2}} 
\begin{pmatrix}
1 & 0 \\
0 & 1
\end{pmatrix}_A \right. \\
&\quad + \left. \begin{pmatrix}
0 \\
\sqrt{\frac{1}{2}}
\end{pmatrix} \otimes \sqrt{\frac{1}{2}} 
\begin{pmatrix}
0 & -1 \\
1 & 0
\end{pmatrix}_A \right),
\end{aligned}
\end{equation}
\begin{equation}
\begin{aligned}
K_{X} &= \left( \begin{pmatrix}
0 \\
\sqrt{\frac{1}{2}}
\end{pmatrix} \otimes \sqrt{\frac{1}{2}} 
\begin{pmatrix}
1 & 1 \\
1 & -1
\end{pmatrix}_\beta \right. \\
&\quad + \left. \begin{pmatrix}
\sqrt{\frac{1}{2}} \\
0
\end{pmatrix} \otimes \sqrt{\frac{1}{2}} 
\begin{pmatrix}
1 & -1 \\
-1 & -1
\end{pmatrix}_\beta \right),
\end{aligned}
\end{equation}
\begin{equation}
Z_1 = |1\rangle \langle 1|_Z \otimes I_{ABX\tilde{A}\tilde{B}}.
\end{equation}

In a comparable manner the last Kraus operators for \(F'\) include the error syndrome details \(w\):

\begin{equation}
\begin{aligned}
K_{Z, w} &= \left( \begin{pmatrix}
0 \\
\sqrt{\frac{1}{2}}
\end{pmatrix} \otimes \sqrt{\frac{1}{2}} 
\begin{pmatrix}
1 & 0 \\
0 & 1
\end{pmatrix}_A \right. \\
&\quad + \left. \begin{pmatrix}
0 \\
\sqrt{\frac{1}{2}}
\end{pmatrix} \otimes \sqrt{\frac{1}{2}} 
\begin{pmatrix}
0 & -1 \\
1 & 0
\end{pmatrix}_A \right) \otimes |w\rangle_W,
\end{aligned}
\end{equation}
\begin{equation}
\begin{aligned}
K_{X, w} &= \left( \begin{pmatrix}
0 \\
\sqrt{\frac{1}{2}}
\end{pmatrix} \otimes \sqrt{\frac{1}{2}} 
\begin{pmatrix}
1 & 1 \\
1 & -1
\end{pmatrix}_\beta \right. \\
&\quad + \left. \begin{pmatrix}
\sqrt{\frac{1}{2}} \\
0
\end{pmatrix} \otimes \sqrt{\frac{1}{2}} 
\begin{pmatrix}
1 & -1 \\
-1 & -1
\end{pmatrix}_\beta \right) \otimes |w\rangle_W,
\end{aligned}
\end{equation}
\begin{equation}
Z_{1, w} = |1\rangle \langle 1|_Z \otimes I_{ABX\tilde{A}\tilde{B}} \otimes |w\rangle_W.
\end{equation}

This formulation provides a detailed description of the BB84 protocol’s quantum phase, the modeling of misalignment and depolarization, and the construction of Kraus operators used in the error correction process.

\subsubsection{Acceptance Test (Parameter Estimation)}

in the acceptance test stage, Alice and Bob publicly share a portion of their random basis choice and a few test bits. Their comparison allows them to approximate the QBER. If the QBER exceeds a predetermined level, the protocol is terminated to ensure security.

For this evaluation, Alice and Bob disclose the measurements and signals for a subset of their data. They decide whether to continue or terminate the protocol based on the computed QBER. The process is modelled with POVMs \(\Gamma_k\) and related expectation values \(\gamma_k\). The characteristics of these POVMs and expectation values differ based on whether fine-graining or coarse-graining is employed within the approval evaluation\cite{wang2022numerical}.

\begin{figure}[ht]
\centering
\includegraphics[width=90mm]{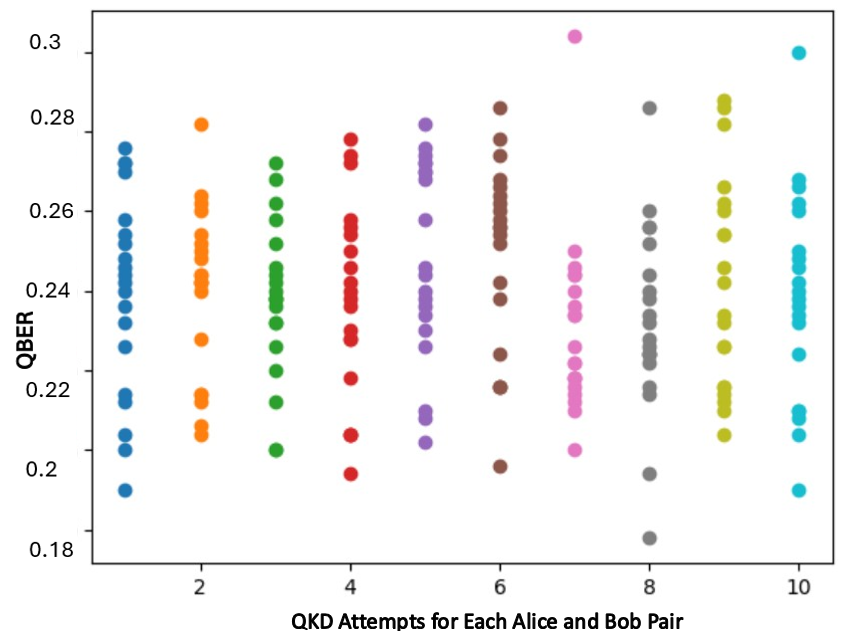}
\caption{QKD Attempts for Each Alice and Bob Pair}
\label{fig:final_key_length}
\end{figure}

To illustrate this, we created a scatter plot of a portion of our dataset, shown in Figure \ref{fig:final_key_length}. The figure presents data from 10 attempts to exchange the secret key between Alice and Bob, assuming 10 pairs of Alice and Bob, each collecting data necessary for training the machine learning model. The full dataset used for training consists of 1,000 by 1,000 or 1,000,000 entries, ensuring comprehensive coverage and robustness in the training data. The dataset includes data transmission rates of 156 Mbps, reflecting real-world conditions as described in \cite{cai2024free}. The scatter plot illustrates the QBER distribution for each Alice and Bob pair.

The goal of the acceptance test is to measure the error rate in the quantum keys exchanged. By announcing their basis choices for a subset of measurements and comparing outcomes, Alice and Bob can compute the QBER, which is crucial for determining the protocol's success. 

Mathematically, this process involves Alice and Bob performing measurements represented by POVMs \(\{\Gamma_k\}\) and obtaining expectation values \(\{\gamma_k\}\). These measurements help estimate the QBER. The QBER calculation is given by:
\begin{equation}
e = \frac{n_{\text{errors}}}{n_{\text{total}}}
\end{equation}
where \(n_{\text{errors}}\) is the number of errors, and \(n_{\text{total}}\) is the total number of qubits compared.

In our approach, we use a machine learning model to predict the final key length based on the number of attempts to generate or share the secret key from the source.

\subsubsection{Classical Processing}
After sifting (keeping only the bits where Alice and Bob used the same basis), Alice and Bob end up with their sifted keys \(X_A\) and \(X_B\).

\subsubsection{Error Correction and Verification}
Both Alice and Bob employ the Cascade protocol for error correction. The Cascade protocol involves multiple rounds of parity checks and binary searches to correct errors in \(X_B\) so that it matches \(X_A\). The Cascade error correction can be represented as:
\begin{equation}
X_B' = \text{Cascade}(X_B, X_A)
\end{equation}
where \(X_B'\) is Bob's corrected key. After error correction, they verify the correctness of their keys using a randomly chosen hash function \(H\):

\begin{equation}
H(X_A) = H(X_B')
\end{equation}
If the hashes match, the protocol continues; otherwise, it aborts.

\subsubsection{Privacy Amplification}
Alice and Bob apply a common two-universal hash function to their raw keys to generate the final secret key:
\begin{equation}
K = h(Z)
\end{equation}
where \(h\) is the hash function and \(K\) is the final secret key.

\subsection{Enhanced method (security aspect)}

The old method for calculating the key rate was considered naive by \cite{tupkary2023using}. They built a new approach, which is discussed in the next subsection.

If \(\tilde{A}\) and \(\tilde{B}\) represent the announcements made by Alice and Bob during blockwise processing, \(Z\) indicates the result of the key map implemented by Alice, and \(E\) denotes Eve's quantum system, then the key rate \(R\) is given by:

\begin{equation}
R = \min_{\rho \in S(\gamma)} S(Z \mid E \tilde{A} \tilde{B}) - p_{\text{pass}} \delta_{\text{leak}}\label{R_incorrect}
\end{equation}
Here:
\begin{itemize}
    \item \(S(Z \mid E \tilde{A} \tilde{B})\) represents the conditional von Neumann entropy of \(Z\) given Eve's information \(E\) and the classical announcements \(\tilde{A}\) and \(\tilde{B}\).
    \item \(p_{\text{pass}}\) denotes the probability that the signal passes sifting.
    \item \(\delta_{\text{leak}}\) is the number of bits used during error correction per bit of raw key.
\end{itemize}
The minimization is over all states \(\rho\) in the set:
\begin{equation}
S(\gamma) = \{ \rho \in \mathcal{H}_+ \mid \text{Tr}(E_k \rho) = \gamma_k \}
\end{equation}
where \(\mathcal{H}_+\) repreents the set of positive semidefinite operators.

\subsection{Cascade Protocol}

The Cascade protocol, a widely recognized method for error correction in QKD, is notable for its simplicity and effectiveness, despite requiring significantly more interactive communication than techniques like low-density parity-check (LDPC) codes, which are computationally intensive due to their iterative decoding processes \cite{xu2020secure}. Understanding the Cascade protocol starts with the BINARY subprotocol, which addresses single errors in bit strings with an odd number of discrepancies and is integral to the error correction process in Cascade.

Initially, if bit strings \(X\) and \(Y\) contain an odd number of errors, Alice splits her string in half and sends the parity of the first half to Bob, who divides his string similarly and announces whether the parity of the first or second half is incorrect. Alice and Bob continue this process on the half with the incorrect parity until Alice identifies the single bit with an error, allowing Bob to correct it. This procedure, requiring the transmission of approximately \(\log(k)\) bits from Alice to Bob and another \(\log(k)\) bits from Bob to Alice, corrects one error in the bit strings \(X\) and \(Y\).

The Cascade protocol involves multiple passes. In the first pass, Alice and Bob partition their bit strings \(X_1 \cdots X_N\) and \(Y_1 \cdots Y_N\), where \(N\) is the total number of sifted bits, into blocks of size \(k_1\). They reveal the parity of each block to detect blocks with an odd number of errors, and for each block with an odd number of errors, they run BINARY to correct one error, ensuring all blocks have an even number of errors by the end of the first pass.

For subsequent passes (\(i \geq 1\)), Alice and Bob select a block size \(k_i\) and a random function \(f_i : [1 \cdots N] \to [1 \cdots N/k_i]\), which assigns each bit to a block in round \(i\). The bits corresponding to positions \(K_{ij} = \{l \mid f_i(l) = j\}\) form the \(j\)-th block in the \(i\)-th round. Alice transmits the parity of each block \(P(A,i,j) = \sum_{l \in K_{ij}} X_l\) to Bob, who computes and announces his parity for the same block. If \(P(A,i,j) \neq P(B,i,j)\), Alice and Bob execute BINARY on the block \(K_{ij}\) to correct one error at position \(l\). This correction identifies multiple error-containing blocks from earlier rounds. Alice and Bob then correct the smallest block from these identified blocks until no blocks have errors, ensuring all blocks in all rounds have an even number of errors by the end of each pass.

A key characteristic of the Cascade protocol is that for every parity bit Alice sends to Bob, Bob reciprocates with the corresponding parity bit, ensuring consistency. Variations of the Cascade protocol exist, differing in block creation methods, block sizes, and the number of passes, yet the core mechanism remains unchanged. Despite requiring communication of the blocks generated in each pass, which are random and independent of the QKD protocol, sharing these details does not provide Eve with any additional information about the key \cite{scarani2009security}.

In essence, the iterative nature and requisite interactive communication of the Cascade protocol allow Alice and Bob to effectively correct errors in their bit strings, making the Cascade protocol a robust error correction method in QKD systems.

\subsection{Problem Definition}

The initial proposal for the Cascade protocol \cite{brassard1993secret} gives a mathematical upper bound \(\delta^{A}_{\text{leak}}\) on the number of bits communicated from Alice to Bob per bit of raw key. This upper bound can also be obtained empirically by running Cascade numerous times with the probable error rate. Regardless of how \(\delta^{A}_{\text{leak}}\) is calculated, we represent the upper bound as \(\delta^{A}_{\text{leak}} = f h(e)\), where \(e\) denotes the error rate in the raw key, \(h(e)\) is the binary entropy function, and \(f\) is the efficiency factor. According to \cite{brassard1993secret, martinez2014demystifying, mao2022high}, typical values for \(f\) range from 1 to 1.5.

In the original Cascade protocol paper \cite{brassard1993secret}, \(\delta^{A}_{\text{leak}}\) is described as the "efficiency" of Cascade, defined as the ratio of the actual number of bits per signal sent from Alice to Bob to \(h(e)\), where \(e\) is the error rate and \(h(e)\) is the binary entropy function. This has led to the mistaken belief that \(\delta^{A}_{\text{leak}}\) represents the true value of \(\delta_{\text{leak}}\) in the key rate calculation when using Cascade in QKD. As a result, the key rate has frequently been erroneously computed using:

\begin{equation}
R_{\text{incorrect}} = \min_{\rho \in S(\gamma)} S(Z \mid E \tilde{A} \tilde{B}) - p_{\text{pass}} \delta^{A}_{\text{leak}},
\end{equation}

where \(R_{\text{incorrect}}\) is the key rate, \(S(Z \mid E \tilde{A} \tilde{B})\) represents the conditional entropy, \(p_{\text{pass}}\) denotes the probability of the protocol passing. This equation fails to account for the communication from Bob to Alice during Cascade. Since Bob’s data are correlated with Alice’s, the communication from Bob can potentially reveal additional information about Alice’s raw key to Eve.

To rectify this oversight, a more accurate formulation that includes the communication from Bob to Alice is necessary \cite{sahu2024state,portmann2022security}. The subsequent sections will delve into the mathematical formulation of this revised key rate calculation. This enhanced method must properly account for all classical communication, ensuring that the security analysis remains rigorous and the key rate computation is accurate. By incorporating the communication from Bob to Alice, we can develop a more comprehensive and precise model for calculating the key rate in QKD protocols utilizing the Cascade protocol.

\subsection{Using Machine Learning to predict Key Rate}

Instead of relying on traditional mathematical models, we utilize an ML technique to predict key rates. In our approach, we specifically employ an autoencoder within the machine learning framework to optimize the performance of QKD systems. The autoencoder is used to efficiently process and predict key rates, such as the final key length, thereby streamlining the error correction process. This integration allows for a more scalable and efficient error correction mechanism, ensuring that the QKD system can handle larger data sizes and higher throughput while maintaining robust security and effective resource management.

The following diagram illustrates the process of autoencoder encoding on Alice's side and decoding on Bob's side, We highlight the prediction of QBER because it is directly related to the key rate in QKD systems. Accurately predicting QBER allows us to estimate the efficiency and security of the key generation process. Since QBER influences the final key length and the overall system's security margin, this prediction is crucial for optimizing resource allocation and enhancing error correction. We rely on the Cascade protocol for error correction, integrating the QBER prediction to more accurately determine the final key length, ensuring the system's scalability and efficiency.

\begin{figure}[h]
    \centering
    \includegraphics[width=\linewidth]{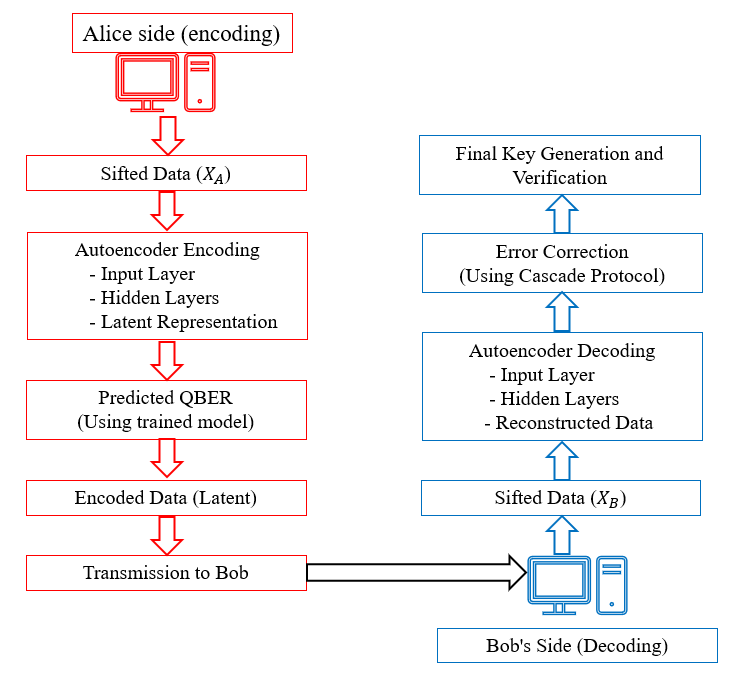} 
    \caption{Diagram of Autoencoder Encoding in Alice and Decoding in Bob for QBER Prediction.}
    \label{fig:autoencoder_diagram}
\end{figure}

The diagram \ref{fig:autoencoder_diagram} illustrates how the autoencoder works on both Alice's and Bob's sides, focusing on predicting the QBER and integrating with the Cascade protocol for error correction. On Alice's side, the process starts with her sifted data, which are the raw key bits after the sifting step where she and Bob have matched their bases. This sifted data is fed into the autoencoder, which compresses it into a compact, latent representation. This latent representation is crucial because it allows Alice to predict the QBER, an essential metric for error correction. Alice then transmits this encoded data to Bob.

On Bob's side, he begins with his sifted data, which corresponds to Alice's sifted data. When Bob receives the latent representation from Alice, he uses the autoencoder's decoding layers to reconstruct the original sifted data. The predicted QBER from Alice's side helps Bob adjust the Cascade protocol parameters for error correction. This ensures that both Alice and Bob's final keys match. After this error correction step, Alice and Bob generate their final secret key. They verify the correctness of their keys using a randomly chosen hash function, ensuring that both keys are identical and secure. This process shows how integrating the autoencoder with traditional QKD methods not only improves efficiency but also enhances the accuracy and reliability of the key exchange.

This framework represents the phases in the QKD protocol using Kraus operators \(\{K_i\}\), which account for the measurements, declarations, and sifting performed by both Alice and Bob and \(\{Z_j\}\), which implement a constricting channel on the key register. So, the requires the computation of \(F\) instead of \(F'\), can be implemented by appropriately modifying the Kraus operators for the optimization issue of \(F'\).

The statistical framework defines the optimization problem for \(F\) as follows:

\begin{equation}
F = \min_{\rho \in S(\gamma)} f(\rho),
\end{equation}

where

\begin{equation}
f(\rho) = D(G(\rho) \parallel Z(G(\rho))),
\end{equation}

\begin{equation}
G(\rho) = \sum_{i} K_i \rho K_i^{\dagger},
\end{equation}

\begin{equation}
Z(G(\rho)) = \sum_{j} Z_j G(\rho) Z_j^{\dagger}.
\end{equation}

The quantum relative entropy is defined as \(D(X \parallel Y) = \text{Tr}(X \log(X)) - \text{Tr}(X \log(Y))\), where \(\log\) represents the matrix logarithm.

To compute \(F'\), an additional announcement declaring \(w = x \oplus y\) must be included. This is done through:

\begin{align}
K_{\alpha, \beta, w} &= \sum_{x, y} \delta_{w, x \oplus y} |r(\alpha, \beta, x)\rangle \langle x|_X \langle y|_Y \langle \alpha, \beta|_{\tilde{A} \tilde{B}} \nonumber \\
&\quad \otimes P_A(\alpha, x) \otimes P_B(\beta, y)
\end{align}

The operators forming the new \(G\) map are \(\{K_i\} = \{K_{\alpha, \beta, w} \mid (\alpha, \beta) \in A\}\). The \(Z\) map is implemented by the operators \(\{Z_i\}\), defined as \(Z_i = |i\rangle \langle i|_Z \otimes I_{ABXY\tilde{A}\tilde{B}W}\).

Consider \(\alpha\) and \(\beta\) as the basis choice announcements made by Alice and Bob. The POVMs for Alice and Bob are \(P_A = \{P_A(\alpha, x)\}\) and \(P_B = \{P_B(\beta, y)\}\), with \(x\) and \(y\) representing measurement outcomes. The announcements \((\alpha, \beta)\) retained post-sifting form the set \(A\), and \(r(\alpha, \beta, x)\) is the key map used by Alice. The Kraus operators for Eq. (7) are:

\begin{align}
K_{\alpha, \beta} = \sum_{x, y} |r(\alpha, \beta, x)\rangle \langle x|_X \langle y|_Y \langle \alpha, \beta|_{\tilde{A} \tilde{B}} \\
\otimes P_A(\alpha, x) \otimes P_B(\beta, y),
\end{align}

The operators generating the \(G\) map are \(\{K_i\} = \{K_{\alpha, \beta} \mid (\alpha, \beta) \in A\}\). The \(Z\) map is carried out by \(\{Z_i\}\), where \(Z_i = |i\rangle \langle i|_Z \otimes I_{ABXY\tilde{A}\tilde{B}}\). The output state \(G(\rho)\) is classical in \(\alpha\) and \(\beta\), indicating that the basis choices are public knowledge to Eve.

For the rest of this paper, we will calculate both \(F = \min_{\rho \in S(\gamma)} f(\rho)\) and \(F' = \min_{\rho \in S(\gamma)} f'(\rho)\) for different BB84 protocol implementations. If \(F = F'\), it suggests that the previous analysis of Cascade was flawed but still produced correct results, leading to identical key rates from Eqs. (2) and (4). If \(F > F'\), it indicates an erroneous previous analysis that gave inaccurate results. The key rate difference between Eqs. (2) and (4) is \(F - F'\), representing the difference in secure key bits per signal. In terms of secure key bits per second, the rate difference is \(R_r (F - F')\), where \(R_r\) is the repetition rate.

This approach assumes that Alice and Bob generate bit strings from their measurements (e.g., \(x\), \(y\), and \(x \oplus y\) are bits). Events like no-detection must either be discarded during sifting or mapped to bits. This assumption is essential for using Cascade, which corrects errors in bit strings. Moreover, since many finite-size key rate analyses optimize the same objective function (\(F\)) over different constraints, our solution can be easily adapted by substituting \(F\) with \(F'\).

\subsection{Mathematical Framework}

In traditional approaches, the key rate \( R \) is often calculated using complex mathematical models that are computationally intensive \cite{winick2018reliable, hu2022robust}. By integrating machine learning, specifically using an autoencoder, we can simplify this process. The autoencoder learns the intrinsic relationships between the system parameters—such as signal-to-noise ratio (SNR), Quantum Bit Error Rate (QBER), and polarization states—and the key length, allowing for a more efficient optimization process.

The optimization problem can be formulated as follows:

\begin{equation}
\text{maximize} \quad r
\end{equation}

subject to

\begin{equation}
k = g(p, q, s) \leq k_{\text{max}},
\end{equation}

where \( k_{\text{max}} \) is the maximum predictable key length, and \( g(p, q, s) \) represents the key length as a function of system parameters such as SNR, QBER, and polarization states.

To solve this optimization problem, we adopt a gradient ascent methodology combined with constraint handling. At each iteration, the parameters \( p \), \( q \), and \( s \) are updated according to:

\begin{equation}
p_{i+1} = p_i + \alpha \frac{\partial g}{\partial p},
\end{equation}

\begin{equation}
q_{i+1} = q_i + \alpha \frac{\partial g}{\partial q},
\end{equation}

\begin{equation}
s_{i+1} = s_i + \alpha \frac{\partial g}{\partial s},
\end{equation}

Here, \( \alpha \) is the learning rate. The partial derivatives \(\frac{\partial g}{\partial p}\), \(\frac{\partial g}{\partial q}\), and \(\frac{\partial g}{\partial s}\) represent how changes in the system parameters affect the predicted key length. These derivatives are learned by the ML model during training.

The autoencoder framework in MATLAB can effectively capture these relationships by minimizing the reconstruction error between the predicted and actual key lengths. This framework provides a robust and efficient means to optimize the key rate, making the process faster and more reliable compared to traditional mathematical models. By leveraging the predictive power of the ML model, we enhance the overall efficiency and accuracy of the QKD system, ensuring that the key rate is maximized while maintaining the necessary security constraints.

\subsection{System Architecture Overview}

To clearly understand the system architecture, we present a high-level diagram that illustrates the integration of the autoencoder and the Cascade protocol within the QKD system.

\begin{figure}[h]
    \centering
    \includegraphics[width=\linewidth]{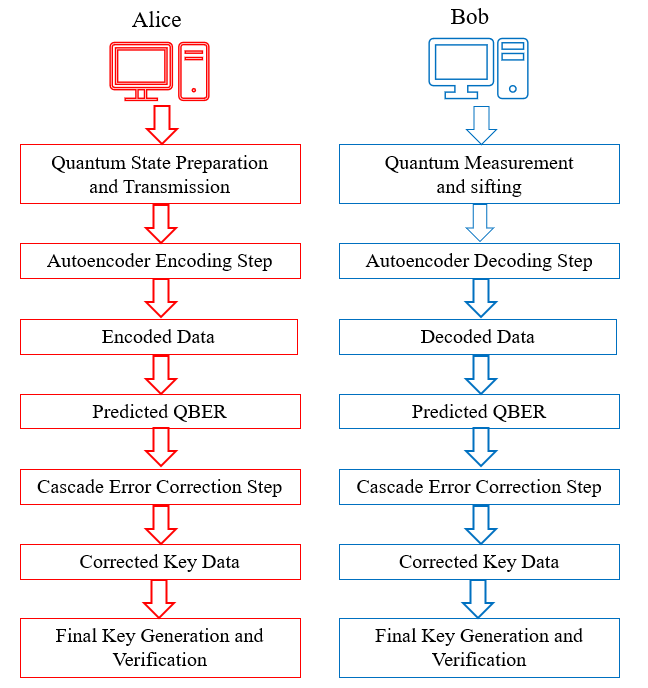} % Replace with the path to your actual image
    \caption{High-level system architecture diagram of the QKD system integrating autoencoder and Cascade protocol.}
    \label{fig:system_architecture}
\end{figure}

As shown in Figure \ref{fig:system_architecture}, the system consists of the following components:

\begin{enumerate}
    \item \textbf{Quantum State Preparation and Transmission (Alice to Bob):} Alice prepares quantum states, typically photons polarized in specific states, and transmits them to Bob. This is the foundational step where quantum keys are generated.
    \item \textbf{Measurement and Sifting Process:} Bob measures the received quantum states. Both Alice and Bob then perform a sifting process, where they compare the bases they used and retain only the bits where their bases match. This ensures that they have correlated data to work with.
    \item \textbf{Autoencoder Encoding and Decoding Steps:} The sifted data, which is now a binary sequence, is fed into an autoencoder. The autoencoder compresses this data into a latent representation, capturing essential features while reducing dimensionality.
    \item \textbf{Prediction of QBER or Key Length:} The autoencoder is used to predict the Quantum Bit Error Rate (QBER) or the final key length. During training, the autoencoder learns the relationship between the input data and these metrics, enabling it to make accurate predictions for new data.
    \item \textbf{Cascade Protocol for Error Correction:} The predicted QBER guides the application of the Cascade protocol. This protocol involves multiple rounds of error correction through parity checks and binary searches to ensure the final keys of Alice and Bob are identical.
    \item \textbf{Final Key Generation and Verification:} After error correction, Alice and Bob generate the final secret key. They then verify the correctness of their keys using a randomly chosen hash function to ensure the keys match, thus confirming the integrity of the key exchange.
\end{enumerate}

\section{Experimental Setup}

This section details the experimental setup used to evaluate the performance of the autoencoder in predicting the QBER and key length in QKD systems. We describe the architecture of the autoencoder, the dataset used, the training procedure, and the evaluation metrics.  The simulations were conducted using MATLAB on a computer with the following specifications: AMD Ryzen 5 5500U with Radeon Graphics at 2.10 GHz, 8.00 GB RAM (7.37 GB usable), and a 64-bit operating system, x64-based processor.

We chose to use an autoencoder for our QKD system due to several key advantages over traditional methods. Autoencoders are particularly adept at managing and learning from high-dimensional data, making them ideal for handling complex and voluminous data generated from quantum states, error rates, and key lengths \cite{mahmud2020variational}. Their ability to compress and reconstruct data efficiently ensures higher predictive accuracy for QBER and final key length, which is crucial for optimizing error correction and robust key generation \cite{mertz2013quantum}. Additionally, autoencoders offer scalability, efficiently processing large datasets required by high-throughput QKD systems, and adaptability, as they can be retrained with new data to accommodate varying noise levels and error rates without extensive recalibration. This adaptability ensures ongoing efficiency and accuracy in dynamic quantum communication environments \cite{zhao2019variational}. Furthermore, autoencoders reduce computational complexity by leveraging neural network operations that can be parallelized and optimized, resulting in quicker convergence and lower overall computation times compared to traditional methods \cite{charte2021reducing}. Integrating the autoencoder with the Cascade protocol for error correction further enhances the system's performance, reducing the number of iterations needed for error correction and leading to faster, more reliable key generation. These benefits collectively make the autoencoder a robust and practical choice for secure quantum communication, addressing the limitations of traditional methods.

\subsection{Dataset}
In this study, we generated our dataset using the simulation algorithms described in \cite{al2022new} and \cite{al2021detecting}. The algorithm employed in this study is both straightforward and highly effective, providing final key lengths and QBER that closely approximate real-world values.  This results in a more accurate and practical representation of QKD operations, enhancing the reliability of the machine learning predictions. Furthermore, this algorithm has been successfully used in other machine learning applications, such as detecting attackers in quantum communication networks, demonstrating its versatility and robustness in enhancing the efficiency and accuracy of various ML techniques \cite{al2021detecting, al2021use, al2021machine}.

Our dataset comprises 1,000,000 instances, each representing an attempt to generate a secret key. These instances cover a wide range of photon transmission rates, from 1 kbps to 156 Mbps, capturing the variability and complexity typical of actual QKD scenarios. For each instance, we recorded essential details such as the initial data, the number of photons transmitted, the QBER, and the resulting final key length.

To ensure the reliability and robustness of our autoencoder model, we split the dataset into training and testing sets. We allocated 80\% of the data for training, enabling the model to learn the underlying patterns and relationships, while the remaining 20\% was reserved for testing.

\subsection{Autoencoder Architecture}

The autoencoder used in our Quantum Key Distribution (QKD) system is designed to predict the key rate. An autoencoder is a type of artificial neural network that is trained to learn efficient representations of data, typically for the purpose of dimensionality reduction or feature learning. In our implementation, the autoencoder plays a crucial role in optimizing the QKD process by accurately predicting key metrics.

The architecture consists of an encoder and a decoder. The encoder maps the input data to a latent representation, which captures the most significant features in a reduced dimensionality. This compressed representation is then used by the decoder to reconstruct the original input data. The goal of the training process is to minimize the reconstruction error, which is the difference between the input data and the reconstructed output.

\begin{figure}[ht]
    \centering
    \includegraphics[width=0.9\linewidth]{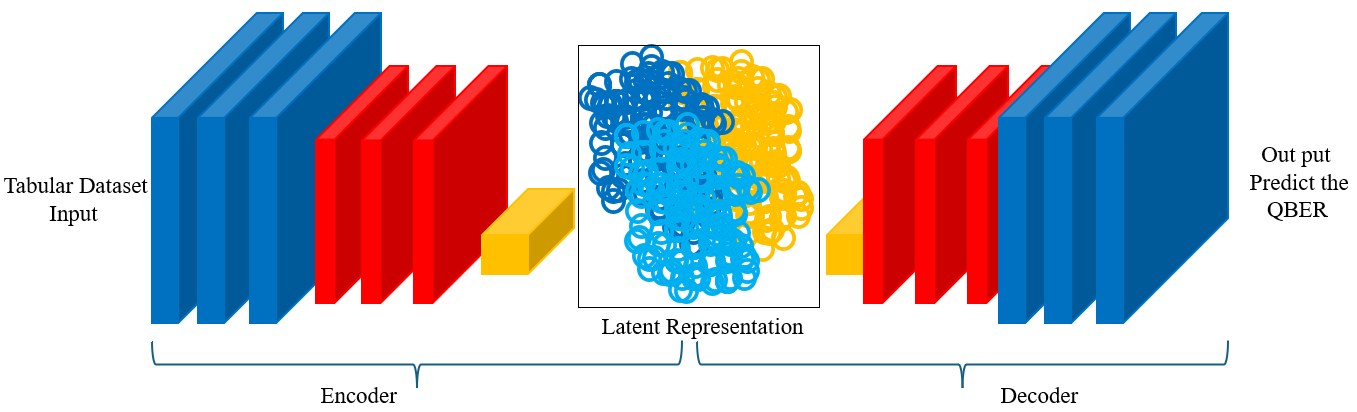}
    \caption{Illustration of the autoencoder architecture used in our QKD system. The encoder compresses the input data into a latent representation, and the decoder reconstructs the original data from this representation.}
    \label{fig:autoencoder_architecture}
\end{figure}

As depicted in Figure \ref{fig:autoencoder_architecture}, the encoder and decoder each consist of multiple layers. The encoder compresses the input data into a latent representation, shown in the middle of the diagram. This representation captures the essential features of the data in a lower-dimensional space. The decoder then reconstructs the data from this latent representation, aiming to match the original input as closely as possible.

For our experiments, we implemented the autoencoder using MATLAB. The encoder comprises multiple layers, each responsible for progressively compressing the input data. The decoder, on the other hand, mirrors this structure and aims to reconstruct the data from the latent representation. Specifically, we used four hidden layers in the autoencoder, which allows it to capture complex patterns in the data.

To train the autoencoder, we used a dataset generated by simulating the QKD process. The dataset included information about the initial photons transmitted and the resulting key lengths after multiple attempts. By training on this dataset, the autoencoder learns to predict the QBER and the final key length accurately. The training process involves minimizing the mean squared error (MSE) between the actual key lengths and the predicted values.

Overall, the autoencoder architecture in our QKD system significantly enhances the accuracy of predicting key metrics, thereby improving the efficiency and security of the quantum communication process.

\subsection{Training Procedure}

In determining the upper limit of bits communicated from Alice to Bob per raw key bit, denoted as $\delta^{A}_{\text{leak}}$, our methodology integrates both theoretical and empirical approaches as outlined by Brassard et al. \cite{brassard1993secret}. This combination of approaches ensures robustness against real-world variations in quantum key distribution (QKD) scenarios.

\subsubsection{Analytical Calculation}
The theoretical upper bound of $\delta^{A}_{\text{leak}}$ is calculated using the binary entropy function $h(e)$, where $e$ represents the error rate in the raw key. The efficiency factor $f$ typically ranges from 1 to 1.5, reflecting the protocol's efficiency under ideal conditions. The calculation is given by:
\begin{equation}
\delta^{A}_{\text{leak}} = f \cdot h(e)
\end{equation}

\subsubsection{Empirical Evaluation}
Empirically, $\delta^{A}_{\text{leak}}$ is derived through multiple simulations of the Cascade protocol under a spectrum of expected error rates. This approach, which has been previously performed in studies such as in \cite{calver2011empirical} and \cite{timothy2011emprical}, allows for refining the efficiency factor based on observed discrepancies:
\begin{itemize}
    \item \textbf{Simulation Runs:} Multiple iterations of Cascade are performed at varied expected error rates to empirically determine $\delta^{A}_{\text{leak}}$.
    \item \textbf{Parameter Adjustment:} Protocol parameters are dynamically adjusted based on the deviations observed between predicted and actual error rates during these simulations.
\end{itemize}

\subsubsection{Iterative Refinement}
This iterative process aligns theoretical predictions with empirical observations, ensuring the calculated $\delta^{A}_{\text{leak}}$ adequately reflects practical QKD implementation scenarios. The continuous adjustment of the efficiency factor and the subsequent recalculations of $\delta^{A}_{\text{leak}}$ optimize the protocol for enhanced security and efficacy in operational environments.

\subsection{Evaluation Metrics}

The Cascade protocol's efficiency is initially assessed using an analytical upper bound \(\delta^{A}_{\text{leak}}\) on the number of bits transmitted from Alice to Bob per bit of raw key, as suggested by Brassard et al. \cite{brassard1993secret}. In real-world applications, this upper bound is typically determined empirically by executing multiple iterations of the Cascade protocol at the expected error rate. Regardless of the method employed to find \(\delta^{A}_{\text{leak}}\), it is commonly expressed as \(\delta^{A}_{\text{leak}} = f h(e)\) \cite{brassard1993secret,martinez2014demystifying,mao2022high}.

In this context, all classical communication is presumed to be known to Eve. However, this key rate expression fails to consider the communication from Bob to Alice during the Cascade protocol, which could potentially reveal additional information about Alice's raw key to Eve. Consequently, a more accurate formulation is required, one that includes the communication from Bob to Alice to ensure the security analysis is rigorous and the key rate computation is precise \cite{tupkary2023using}.

In our approach, we employ machine learning techniques, specifically an autoencoder, to predict key rates. The autoencoder framework implemented in MATLAB captures the relationships between system parameters and the key length, simplifying the optimization process. This framework replaces traditional mathematical models with a more efficient machine learning approach, which is especially beneficial for complex and dynamic environments in QKD systems.

The autoencoder framework is evaluated based on its ability to minimize the reconstruction error between predicted and actual key lengths. The mean squared error (MSE) is used as the primary metric for assessing the accuracy of the model:

\begin{equation}
MSE = \frac{1}{n} \sum_{i=1}^{n} (y_i - \hat{y}_i)^2
\end{equation}

where \(n\) is the number of data points, \(y_i\) is the true value, and \(\hat{y}_i\) is the predicted value. The effectiveness of the autoencoder and its integration with the Cascade protocol is demonstrated through training loss values at various epochs, showcasing the model's capability to accurately predict the Quantum Bit Error Rate (QBER) before transmitting the quantum key. As the model training progresses, the decreasing loss values indicate improved error correction and a lower QBER, thereby enhancing the security and efficiency of the QKD system.

\begin{table}[ht]
\centering
\caption{Training Loss at Specific Epochs}
\label{Training loss at specific epochs.}
\begin{tabular}{|c|c|}
\hline
Epoch & Loss \\
\hline
1     & 0.4565 \\
13    & 0.2240 \\
50    & 0.0913 \\
70    & 0.0405 \\
100   & 0.0194 \\
\hline
\end{tabular}
\end{table}

Table \ref{Training loss at specific epochs.} shows the training loss at the 1st, 13th, 50th, 70th, and 100th epochs, illustrating the model's learning process. The table indicates a clear trend of decreasing loss values, reflecting the model's increasing accuracy and robustness in predicting QBER and enhancing the overall performance of the QKD system.

By continually improving the error correction capability, the iterative training process of our integrated model significantly boosts both the efficiency and security of our QKD system, making it a promising solution for real-world quantum communication applications.

To evaluate the simulation, we used the following standard evaluation metrics: accuracy, and MSE.
\begin{equation}
\text{Accuracy} = \frac{\text{Number of correct predictions}}{\text{Total number of predictions}}
\end{equation}

\subsection{Algorithm Implementation}

The implementation of the autoencoder training with the Cascade protocol for error correction is described in Algorithm \ref{alg:autoencoder_cascade}. The algorithm begins by initializing the decoded output, represented as a set of values $\{x_i\}_{i=1}^{N}$. It then applies the Cascade protocol to this output to correct any errors, resulting in the \texttt{corrected\_output}.

\begin{algorithm}[]
  \caption{Autoencoder Training with Cascade Protocol Error Correction}
  \label{alg:autoencoder_cascade}
  
  \SetKwInOut{Input}{Input}
  \SetKwInOut{Output}{Output}
  
  \Input{Decoded output $\{x_i\}_{i=1}^{N}$}
  \Output{Final output $\text{final\_output}$}
  
  Apply Cascade protocol to $\{x_i\}_{i=1}^{N}$ to obtain $\{x_i'\}_{i=1}^{N}$\;
  Define autoencoder neural network $f_{\theta}$ with parameters $\theta$\;
  Initialize Adam optimizer with learning rate $\eta$\;
  Define training operation for autoencoder\;
  
  \For{each training epoch $t$}{
    Randomly select mini-batch indices $\mathcal{I}$ from $\{x_i'\}_{i=1}^{N}$\;
    Extract $\text{batch\_input}$ and $\text{batch\_target}$ from $\{x_i'\}_{i=1}^{N}$ using $\mathcal{I}$\;
    Compute predicted output $\hat{y} = f_{\theta}(\text{batch\_input})$\;
    Compute loss $L = \frac{1}{|I|}\sum_{i \in I}(y_i - f_{\theta}(x_i))^2$\;
    Compute gradients $\nabla_{\theta} L$\;
    Update $\theta \leftarrow \theta - \eta \nabla_{\theta} L$\;
  }
  
  Apply $f_{\theta}$ to $\{x_i'\}_{i=1}^{N}$ to obtain $\text{final\_output}$\;
\end{algorithm}

Algorithm \ref{alg:autoencoder_cascade} begins by taking the decoded output $\{x_i\}_{i=1}^{N}$ and applying the Cascade protocol to correct any errors, resulting in the \texttt{corrected\_output}:

\begin{equation}
\text{corrected}_{\text{output}} = \text{CascadeProtocol}_{\text{output}}(\text{decoded}_{\text{output}})
\end{equation}

Next, an autoencoder neural network, $f_{\theta}$ with parameters $\theta$, is introduced to this corrected output. The autoencoder is represented by the equation:

\begin{equation}
f_{\theta}(x) = \theta^T x
\end{equation}

The Adam optimizer \cite{kingma2014adam}, with a predefined learning rate $\eta$, is used to adaptively update the network's weights during training. For each training epoch, a random mini-batch of indices $I$ is selected from the \texttt{corrected\_output}. Corresponding \texttt{batch\_input} and \texttt{batch\_target} are derived using these indices. The autoencoder model is applied to the \texttt{batch\_input}, producing a predicted output $\hat{y}$:

\begin{equation}
\hat{y} = f_{\theta}(\text{batch\_input})
\end{equation}

The mean squared error loss $L$ is then computed between the predicted output $\hat{y}$ and the \texttt{batch\_target}:

\begin{equation}
L = \frac{1}{|I|}\sum_{i \in I}(y_i - f_{\theta}(x_i))^2
\end{equation}

Gradients of this loss with respect to the trainable parameters $\theta$ are calculated:

\begin{equation}
\nabla L = \frac{\partial L}{\partial \theta}
\end{equation}

These gradients are used to update the parameters via the Adam optimizer:

\begin{equation}
\theta_{\text{new}} = \text{AdamOptimizer}(\theta_{\text{old}}, \nabla L, \eta)
\end{equation}

This process is repeated for a predefined number of epochs. Finally, after the training phase, the autoencoder model is applied to the reshaped \texttt{corrected\_output} to yield the \texttt{final\_output}:

\begin{equation}
\text{final\_output} = f_{\theta}(\text{corrected\_output})
\end{equation}

It should be noted that this description simplifies the full procedure.

Table \ref{Training loss at specific epochs.} shows the training loss of our integrated model at the 1st, 13th, 50th, 70th, and 100th epochs. This subset of the total training epochs illustrates a trend where the loss decreases as training progresses, indicating the effectiveness of the learning process in our model, which comprises the autoencoder and the Cascade protocol.

The decreasing loss values suggest that the model becomes increasingly capable of accurately predicting QBER before transmitting a quantum key. This is crucial because it demonstrates the effectiveness of the Cascade protocol's iterative error correction methods incorporated into our model. As the loss values decrease, the predictions become more accurate, leading to a lower QBER.

As we approach the later epochs, the model fine-tunes itself better, becoming more resilient against errors. This improvement, depicted by reducing the loss values, results in a more accurate forecast of the QBER.

This progress in model training is beneficial for our QKD system. A lower error in QBER prediction or estimation will provide a better indication of Eve's presence, hence, enhanced security. Therefore, as the table suggests, the iterative training process significantly improves both the efficiency and security of our QKD system by continually improving the error correction capability.

%%%%%%%%%%%%%%%%%%%%%%%%%%%%%%%%%%%
%%%%%%%%%%%%%%%%%%%%%%%%%%%%%%%%%%%%
%%%%%%%%%%%%%%%%%%%%%%%%%%%%%%%%%%%%

\section{Results and Discussion}

To demonstrate the improved efficiency of using an autoencoder for predicting the final key length and the QBER in QKD compared to traditional mathematical methods, we analyze and compare the computational complexity and time efficiency of both approaches.

\subsection{Computational Complexity of Traditional Mathematical Methods}

Traditional mathematical models for QKD typically involve the following steps:
\begin{itemize}
    \item \textbf{State Preparation and Measurement}: Represented by preparing and measuring quantum states.
    \item \textbf{Parameter Estimation}: Estimating parameters such as error rates.
    \item \textbf{Key Rate Calculation}: Calculating the key rate using complex formulas.
\end{itemize}

Let the computational complexity of these steps be:
\begin{itemize}
    \item \textbf{State Preparation and Measurement}: \(O(n)\)
    \item \textbf{Parameter Estimation}: \(O(n^2)\)
    \item \textbf{Key Rate Calculation}: \(O(n^3)\)
\end{itemize}

Thus, the overall complexity can be approximated as:
\[
T_{\text{traditional}} = O(n) + O(n^2) + O(n^3) = O(n^3)
\]

\subsection{Computational Complexity of Using Autoencoders}

The autoencoder approach involves the following operations:
\begin{itemize}
    \item \textbf{Data Encoding}: Encoding the input data into a latent representation.
    \item \textbf{Data Decoding}: Reconstructing the input data from the latent representation.
    \item \textbf{Error Correction with Cascade Protocol}: Correcting errors in the decoded data.
\end{itemize}

The complexity of these steps can be described as:
\begin{itemize}
    \item \textbf{Data Encoding}: \(O(n \cdot h)\), where \(n\) is the input size and \(h\) is the size of the hidden layer.
    \item \textbf{Data Decoding}: \(O(h \cdot n)\)
    \item \textbf{Error Correction with Cascade Protocol}: \(O(n \log n)\), as Cascade typically involves logarithmic iterations for error correction.
\end{itemize}

Thus, the overall complexity of the autoencoder approach is:
\[
T_{\text{autoencoder}} = O(n \cdot h) + O(h \cdot n) + O(n \log n)
\]

Given that \(h \ll n\) (hidden layer size is much smaller than input size), this simplifies to:
\[
T_{\text{autoencoder}} \approx O(n \log n)
\]

\subsection{Comparison of Computational Complexities}

To compare the time efficiencies, we examine the order of growth of both approaches:
\begin{itemize}
    \item \textbf{Traditional Method}: \(O(n^3)\)
    \item \textbf{Autoencoder Method}: \(O(n \log n)\)
\end{itemize}

Since \(O(n \log n) \ll O(n^3)\) for large \(n\), the autoencoder approach is significantly faster than the traditional method.

\subsection{ML Accuracy}

ML model in predicting the QBER and final key length for QKD is a crucial factor in the overall efficiency and security of the system. The ML model's prediction accuracy ensures that error correction and key generation processes are both robust and efficient.
\begin{figure}[!t]
\centering
\includegraphics[width=88mm,height=70mm]{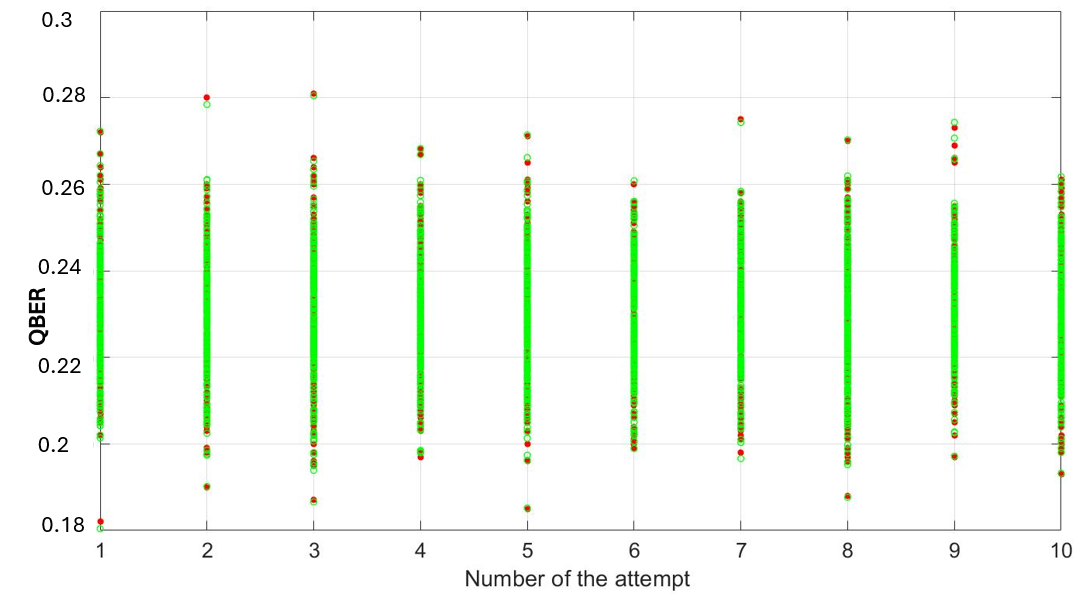}
\caption{Prediction of key length across 10 trials using an initial photon input of 1 kbps}
\label{fig1}
\end{figure}

Figure \ref{fig1} shows a sample of 10 attempts to generate a QKD using a fixed initial number of photons. Each trial includes 10 attempts to generate a new QKD. The figure illustrates the number of final key lengths obtained from each trial. This sample from the actual data demonstrates the impact of ML protection, particularly in smaller datasets. The prediction accuracy of the model is represented by the green color, while the red color explicitly indicates when it falls outside the threshold bounds. The model achieved an accuracy rate of more than 99\% in predicting QBER, and the MSE of the model was 17\%.

\subsection{Time consumption}

The chosen input size of 156 Mbps reflects real-world scenarios where high data rates are common, especially in modern quantum key distribution (QKD) systems. High data rates are essential for practical and scalable QKD applications, making it crucial to evaluate the performance of both traditional and ML-based methods under these conditions.

Evaluating the performance for large input sizes allows us to understand the scalability of both approaches. While traditional methods may perform adequately for smaller input sizes, their computational complexity becomes prohibitive as input sizes grow. Conversely, ML-based methods are designed to handle large datasets efficiently, making them more suitable for high-throughput QKD systems.

\begin{figure}[!t]
\centering
\includegraphics[width=95mm,height=70mm]{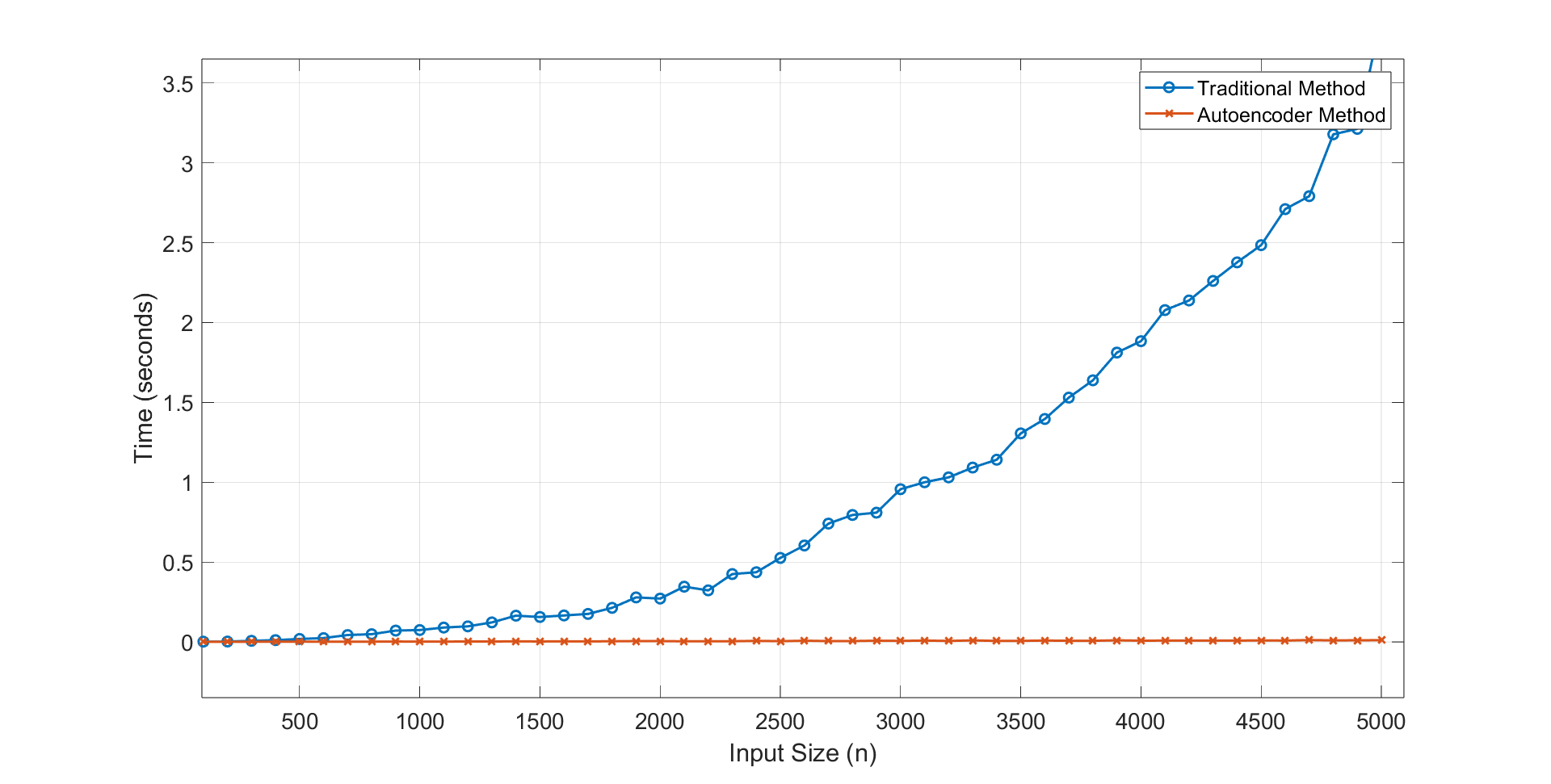}
\caption{Comparison of Computation Time: Traditional Method vs Autoencoder Method}
\label{fig:comparison}
\end{figure}

Figure \ref{fig:comparison} shows the comparison of computation time between the traditional method and the autoencoder method. As depicted, the traditional method's computation time increases significantly with larger input sizes, while the autoencoder method maintains a relatively constant and low computation time. This highlights the efficiency of the ML-based approach in handling large datasets, proving its superiority over traditional methods for scalable QKD implementations.

To further validate these findings, figure 7 will present results with even larger datasets, specifically 1.56, 15.6, and 156 Mbps, to demonstrate the realistic performance differences in more demanding scenarios.

During our initial attempt to compare the computational efficiency of traditional and autoencoder-based methods for Quantum Key Distribution (QKD), we encountered memory limitations due to the large input size of 156 Mbps. The traditional method's operations, particularly matrix multiplications, exceeded MATLAB's maximum array size preference. To address this issue, we revised our approach to simulate the computational time for each step instead of performing the actual operations. This simulation approach accurately reflects the computational complexity of each method, allowing us to effectively compare their efficiencies without running into memory constraints.

%%%%%%%%%%%%

\begin{figure}[!t]
\centering
\includegraphics[width=95mm,height=70mm]{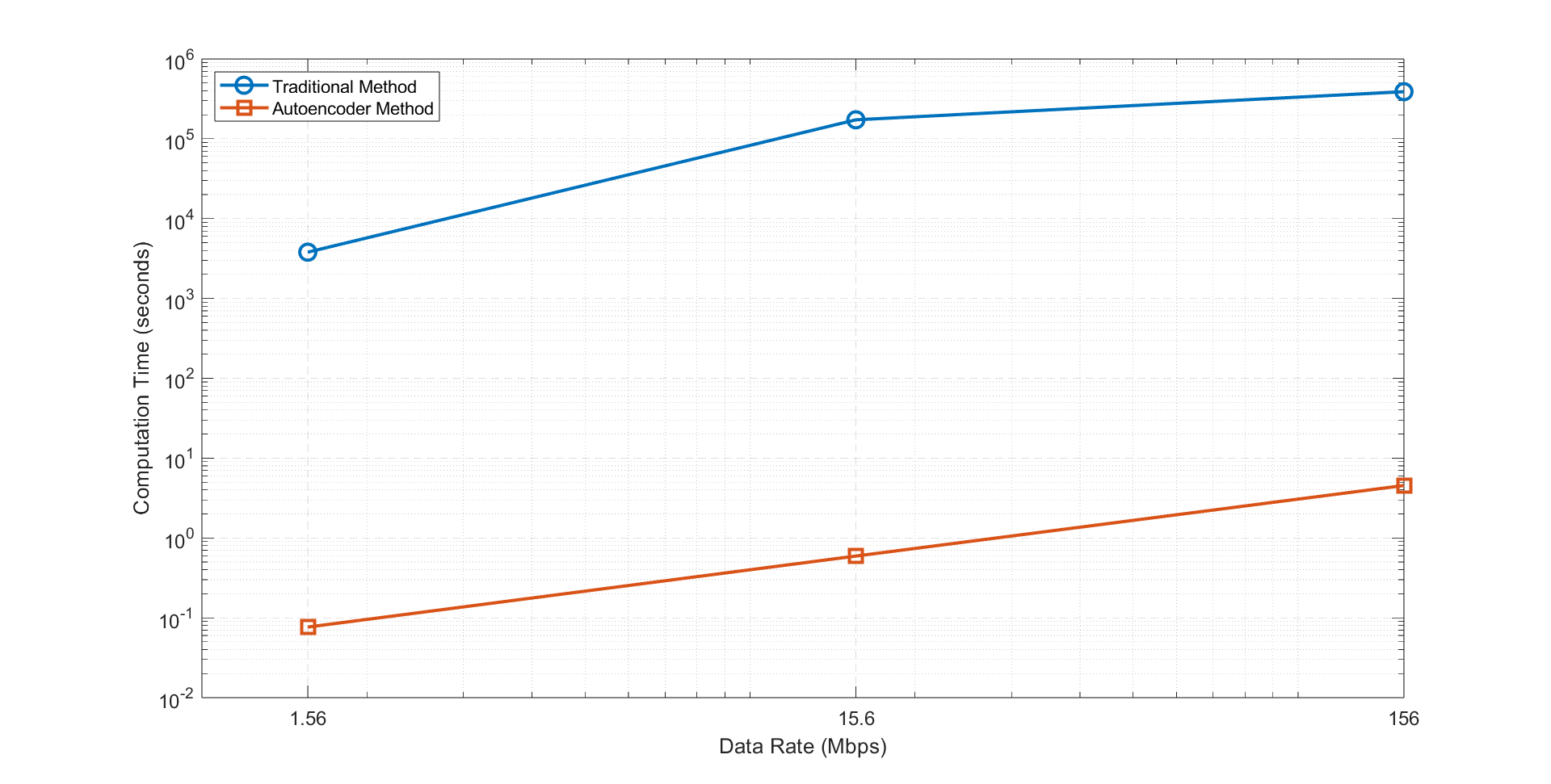}
\caption{Computation Time Comparison: Traditional Method vs. Autoencoder Method (Real Data)}
\label{fig:comparison_big}
\end{figure}

Figure \ref{fig:comparison_big} illustrates the comparison of computation times between the traditional method and the autoencoder method for Quantum Key Distribution (QKD) at different data rates: 1.56 Mbps, 15.6 Mbps, and 156 Mbps. The logarithmic line plot demonstrates a stark contrast in performance between the two methods. The traditional method exhibits significantly higher computation times, especially as the data rate increases, with times ranging from approximately 3,798.9 seconds (about 1.05 hours) for 1.56 Mbps, to 172,800 seconds (two days) for 15.6 Mbps, and more than 4 days for 156 Mbps. In contrast, the autoencoder method maintains a much lower and relatively stable computation time across all data rates, with times of 0.0765 seconds for 1.56 Mbps, 0.5930 seconds for 15.6 Mbps, and 4.5385 seconds for 156 Mbps. This substantial reduction in computation time highlights the efficiency and scalability of the autoencoder approach, making it a more suitable solution for high-throughput QKD systems. The logarithmic scale effectively captures the exponential difference in performance, underscoring the autoencoder method's ability to handle large datasets efficiently compared to the traditional method. These results are based on real data, demonstrating the practical applicability and significant advantages of the autoencoder method in realistic scenarios.

%%%%%%%%%%%%%

One notable observation from our experiments is that the computation time for the ML-based method (autoencoder) remains relatively constant as the input size increases. This behavior can be attributed to several key factors:

\textbf{Fixed Computational Complexity}: The autoencoder model has a fixed architecture, meaning the number of layers and the size of each layer are predetermined. As a result, the time required for predictions is almost constant, regardless of the input size. The inference step, which involves using the trained model to make predictions, has a lower computational complexity compared to the training phase.

\textbf{Parallelism and Optimization}: Modern machine learning frameworks, such as TensorFlow, PyTorch, and MATLAB's deep learning toolbox, are optimized for parallel computation. These frameworks efficiently utilize advanced hardware, including GPUs and TPUs, which are designed to process large datasets simultaneously. This optimization allows for faster computation times that do not scale linearly with the input size.

\textbf{Batch Processing}: During inference, the autoencoder processes data in batches, handling multiple data points in parallel. This batch processing capability leads to more efficient computations, ensuring that the time required for a single prediction remains consistent even as the total input size increases.

\textbf{Pre-Trained Model}: The heavy computational load associated with training the model has already been completed. The prediction phase, which is being measured in these experiments, is significantly less computationally intensive than the training phase.

\textbf{Efficient Algorithms}: Autoencoders and neural network models employ efficient algorithms for operations like matrix multiplication. These algorithms are optimized to reduce computation time, leveraging advanced mathematical techniques to enhance performance.

Figure \ref{fig:comparison} illustrates the comparison of computation time between the traditional method and the autoencoder method. As shown, the traditional method's computation time increases significantly with larger input sizes, while the autoencoder method maintains a relatively constant and low computation time. This demonstrates the efficiency and scalability of the ML-based approach, making it more suitable for high-throughput QKD systems.

To further validate these findings, subsequent figures will present results with even larger datasets, specifically 1.56, 15.6, and 156 Mbps, to demonstrate the realistic performance differences in more demanding scenarios.

The traditional method involves operations with high computational complexity, such as matrix multiplications and inversions, which scale poorly with input size. This leads to exponential increases in computation time as input sizes grow. In contrast, the ML-based method, particularly using autoencoders, leverages parallel processing and optimized algorithms, resulting in significantly lower computational complexity and faster execution times. Our simulations are designed to accurately reflect the computational complexity of both methods. The observed performance improvements are not merely theoretical but are grounded in the inherent differences in how the two approaches handle large-scale data processing. The traditional method's reliance on large matrix operations is inherently less efficient compared to the ML method's use of streamlined, parallelizable neural network operations. In practical implementations, ML models can be further optimized using techniques such as hardware acceleration (e.g., GPUs, TPUs) and distributed computing, which are not as effective for traditional methods due to their sequential nature. These optimizations further enhance the practicality and efficiency of the ML-based approach in real-world scenarios.

The ML-based approach not only speeds up the process but also enhances error correction efficiency. By accurately predicting the Quantum Bit Error Rate (QBER) and final key length, the ML model ensures robust error correction with fewer iterations, leading to faster convergence and higher overall throughput. This is particularly relevant when integrated with the Cascade protocol, where the ML model's predictions enhance the protocol's iterative error correction capabilities, resulting in a more efficient and effective process. The adaptability of ML models to various data patterns and noise levels in QKD systems further justifies their use. Traditional methods require extensive recalibration and tuning for different conditions, whereas ML models can be retrained or fine-tuned with new data, providing a flexible and practical solution for dynamic environments. The significant performance improvement observed in our simulations is a direct result of the fundamental differences in computational complexity and the inherent advantages of ML-based methods for handling large-scale data. Our evaluation reflects a realistic and practical scenario for modern QKD systems, demonstrating the feasibility and efficiency of adopting ML techniques over traditional methods for key length and QBER protection. We are confident that our approach provides a robust and scalable solution for high-throughput QKD applications, addressing the practical challenges of real-world implementations, see figure \ref{fig:comparison_big}.

The significant difference in computational time between the traditional method and the ML-based method for predicting QBER or final key length can be attributed to the fundamental differences in their computational complexities. The traditional method involves high-complexity operations such as matrix multiplications and inversions, which scale poorly with increasing data size, leading to exponentially longer computation times. For example, when the data size increases, the time taken by the traditional method rises dramatically due to its $O(n^{3})$ complexity. In contrast, the ML-based method, particularly when using an autoencoder, performs inference through efficient linear or near-linear operations that are largely independent of data size once the model is trained. This results in relatively constant and significantly faster computation times. The autoencoder's inference phase benefits from fixed computational costs and the ability to leverage parallel processing hardware, making the ML-based approach highly suitable for real-time and large-scale QKD applications. Consequently, while the traditional method may take several hours or even days to process large datasets, the ML-based method can achieve the same task in a matter of minutes, demonstrating its superior efficiency and practicality for modern quantum communication systems.

\section{Conclusion and Future Directions}

In this paper, we have presented a novel approach to enhancing the efficiency and scalability of quantum key distribution (QKD) systems by integrating machine learning (ML) techniques, specifically an autoencoder, with the Cascade protocol for error correction. Traditional mathematical models used to calculate key rates in QKD are computationally intensive and inefficient, particularly for high data rates required in practical and scalable QKD applications. Our ML-based method addresses these challenges by predicting the Quantum Bit Error Rate (QBER) and final key length with high accuracy, significantly reducing the time and computational resources needed for error correction.

We have demonstrated that the ML-based approach maintains a relatively constant and low computation time even as the input size increases, unlike traditional methods whose computation time grows prohibitively. This makes our method more suitable for high-throughput QKD systems. Our experiments, conducted using real-world data rates of up to 156 Mbps, highlight the superiority of the ML-based method over traditional techniques, proving its scalability and efficiency. Notably, our model achieved an accuracy rate of over 99\% in predicting key length or the QBER, further underscoring its reliability and effectiveness.

Future research could explore and compare the performance of various ML models, such as convolutional neural networks (CNNs) or recurrent neural networks (RNNs), to determine if they offer better accuracy or efficiency in predicting QBER and key lengths. Including decoy states in the QKD protocol could further enhance the security and efficiency of key distribution. Future work could investigate the integration of decoy-state methods with our ML-based approach. Our current model could be extended to include additional variables that may impact QBER and key length, providing a more comprehensive understanding of the QKD system's performance under various conditions.

The success of integrating ML with the Cascade protocol suggests that similar techniques could be applied to other error correction protocols, such as low-density parity-check (LDPC) codes, to improve their performance in QKD systems. While our approach is generally applicable to QKD systems, future research could focus on the other types of the QKD protocols, to further enhance their efficiency and security.

Therefore, the integration of ML techniques with error correction protocols represents a significant advancement in the field of QKD. By leveraging the power of ML, we can achieve more efficient, scalable, and secure quantum communication systems, paving the way for their widespread adoption in practical applications.

%%%%%%%%%%%%%%%%%%%%%%%%%
%%%%%%%%%%%%%%%

\bibliography{references}

\begin{thebibliography}{10}
\providecommand{\url}[1]{#1}
\csname url@samestyle\endcsname
\providecommand{\newblock}{\relax}
\providecommand{\bibinfo}[2]{#2}
\providecommand{\BIBentrySTDinterwordspacing}{\spaceskip=0pt\relax}
\providecommand{\BIBentryALTinterwordstretchfactor}{4}
\providecommand{\BIBentryALTinterwordspacing}{\spaceskip=\fontdimen2\font plus
\BIBentryALTinterwordstretchfactor\fontdimen3\font minus \fontdimen4\font\relax}
\providecommand{\BIBforeignlanguage}[2]{{%
\expandafter\ifx\csname l@#1\endcsname\relax
\typeout{** WARNING: IEEEtran.bst: No hyphenation pattern has been}%
\typeout{** loaded for the language `#1'. Using the pattern for}%
\typeout{** the default language instead.}%
\else
\language=\csname l@#1\endcsname
\fi
#2}}
\providecommand{\BIBdecl}{\relax}
\BIBdecl

\bibitem{nguyen2021security}
V.-L. Nguyen, P.-C. Lin, B.-C. Cheng, R.-H. Hwang, and Y.-D. Lin, ``Security and privacy for 6g: A survey on prospective technologies and challenges,'' \emph{IEEE Communications Surveys \& Tutorials}, vol.~23, no.~4, pp. 2384--2428, 2021.

\bibitem{al2021use}
H.~A. Al-Mohammed and E.~Yaacoub, ``On the use of quantum communications for securing iot devices in the 6g era,'' in \emph{2021 IEEE International Conference on Communications Workshops (ICC Workshops)}.\hskip 1em plus 0.5em minus 0.4em\relax IEEE, 2021, pp. 1--6.

\bibitem{jiang2021road}
W.~Jiang, B.~Han, M.~A. Habibi, and H.~D. Schotten, ``The road towards 6g: A comprehensive survey,'' \emph{IEEE Open Journal of the Communications Society}, vol.~2, pp. 334--366, 2021.

\bibitem{ecker2021strategies}
S.~Ecker, B.~Liu, J.~Handsteiner, M.~Fink, D.~Rauch, F.~Steinlechner, T.~Scheidl, A.~Zeilinger, and R.~Ursin, ``Strategies for achieving high key rates in satellite-based qkd,'' \emph{npj Quantum Information}, vol.~7, no.~1, p.~5, 2021.

\bibitem{chapuran2009optical}
T.~Chapuran, P.~Toliver, N.~Peters, J.~Jackel, M.~Goodman, R.~Runser, S.~McNown, N.~Dallmann, R.~Hughes, K.~McCabe \emph{et~al.}, ``Optical networking for quantum key distribution and quantum communications,'' \emph{New Journal of Physics}, vol.~11, no.~10, p. 105001, 2009.

\bibitem{wang2017long}
L.-J. Wang, K.-H. Zou, W.~Sun, Y.~Mao, Y.-X. Zhu, H.-L. Yin, Q.~Chen, Y.~Zhao, F.~Zhang, T.-Y. Chen \emph{et~al.}, ``Long-distance copropagation of quantum key distribution and terabit classical optical data channels,'' \emph{Physical Review A}, vol.~95, no.~1, p. 012301, 2017.

\bibitem{liu2017experimental}
W.-Y. Liu, X.-F. Zhong, T.~Wu, F.-Z. Li, B.~Jin, Y.~Tang, H.-M. Hu, Z.-P. Li, L.~Zhang, W.-Q. Cai \emph{et~al.}, ``Experimental free-space quantum key distribution with efficient error correction,'' \emph{Optics Express}, vol.~25, no.~10, pp. 10\,716--10\,723, 2017.

\bibitem{tupkary2023using}
D.~Tupkary and N.~L{\"u}tkenhaus, ``Using cascade in quantum key distribution,'' \emph{Physical Review Applied}, vol.~20, no.~6, p. 064040, 2023.

\bibitem{briegel1998quantum}
H.-J. Briegel, W.~D{\"u}r, J.~I. Cirac, and P.~Zoller, ``Quantum repeaters: the role of imperfect local operations in quantum communication,'' \emph{Physical Review Letters}, vol.~81, no.~26, p. 5932, 1998.

\bibitem{dai2020towards}
H.~Dai, Q.~Shen, C.-Z. Wang, S.-L. Li, W.-Y. Liu, W.-Q. Cai, S.-K. Liao, J.-G. Ren, J.~Yin, Y.-A. Chen \emph{et~al.}, ``Towards satellite-based quantum-secure time transfer,'' \emph{Nature Physics}, vol.~16, no.~8, pp. 848--852, 2020.

\bibitem{fiandrino2020machine}
C.~Fiandrino, C.~Zhang, P.~Patras, A.~Banchs, and J.~Widmer, ``A machine-learning-based framework for optimizing the operation of future networks,'' \emph{IEEE Communications Magazine}, vol.~58, no.~6, pp. 20--25, 2020.

\bibitem{zhao2018resource}
Y.~Zhao, Y.~Cao, W.~Wang, H.~Wang, X.~Yu, J.~Zhang, M.~Tornatore, Y.~Wu, and B.~Mukherjee, ``Resource allocation in optical networks secured by quantum key distribution,'' \emph{IEEE Communications Magazine}, vol.~56, no.~8, pp. 130--137, 2018.

\bibitem{cao2022evolution}
Y.~Cao, Y.~Zhao, Q.~Wang, J.~Zhang, S.~X. Ng, and L.~Hanzo, ``The evolution of quantum key distribution networks: On the road to the qinternet,'' \emph{IEEE Communications Surveys \& Tutorials}, vol.~24, no.~2, pp. 839--894, 2022.

\bibitem{al2021machine}
H.~A. Al-Mohammed, A.~Al-Ali, E.~Yaacoub, U.~Qidwai, K.~Abualsaud, S.~Rzewuski, and A.~Flizikowski, ``Machine learning techniques for detecting attackers during quantum key distribution in iot networks with application to railway scenarios,'' \emph{IEEE Access}, vol.~9, pp. 136\,994--137\,004, 2021.

\bibitem{alleaume2009topological}
R.~Alleaume, F.~Roueff, E.~Diamanti, and N.~L{\"u}tkenhaus, ``Topological optimization of quantum key distribution networks,'' \emph{New Journal of Physics}, vol.~11, no.~7, p. 075002, 2009.

\bibitem{wallnofer2020machine}
J.~Walln{\"o}fer, A.~A. Melnikov, W.~D{\"u}r, and H.~J. Briegel, ``Machine learning for long-distance quantum communication,'' \emph{PRX Quantum}, vol.~1, no.~1, p. 010301, 2020.

\bibitem{mao2022high}
H.-K. Mao, Q.~Li, P.-L. Hao, B.~Abd-El-Atty, and A.~M. Iliyasu, ``High performance reconciliation for practical quantum key distribution systems,'' \emph{Optical and Quantum Electronics}, vol.~54, no.~3, p. 163, 2022.

\bibitem{gumucs2021novel}
K.~G{\"u}m{\"u}{\c{s}}, T.~A. Eriksson, M.~Takeoka, M.~Fujiwara, M.~Sasaki, L.~Schmalen, and A.~Alvarado, ``A novel error correction protocol for continuous variable quantum key distribution,'' \emph{Scientific reports}, vol.~11, no.~1, p. 10465, 2021.

\bibitem{bennett2014quantum}
C.~H. Bennett and G.~Brassard, ``Quantum cryptography: Public key distribution and coin tossing,'' \emph{Theoretical computer science}, vol. 560, pp. 7--11, 2014.

\bibitem{ekert1991quantum}
A.~K. Ekert, ``Quantum cryptography based on bell’s theorem,'' \emph{Physical review letters}, vol.~67, no.~6, p. 661, 1991.

\bibitem{hong2016challenges}
K.~W. Hong, O.-M. Foong, and T.~J. Low, ``Challenges in quantum key distribution: A review,'' in \emph{Proceedings of the 4th International Conference on Information and Network Security}, 2016, pp. 29--33.

\bibitem{xu2020secure}
F.~Xu, X.~Ma, Q.~Zhang, H.-K. Lo, and J.-W. Pan, ``Secure quantum key distribution with realistic devices,'' \emph{Reviews of modern physics}, vol.~92, no.~2, p. 025002, 2020.

\bibitem{brassard1993secret}
G.~Brassard and L.~Salvail, ``Secret-key reconciliation by public discussion,'' in \emph{Workshop on the Theory and Application of of Cryptographic Techniques}.\hskip 1em plus 0.5em minus 0.4em\relax Springer, 1993, pp. 410--423.

\bibitem{ding2020predicting}
H.-J. Ding, J.-Y. Liu, C.-M. Zhang, and Q.~Wang, ``Predicting optimal parameters with random forest for quantum key distribution,'' \emph{Quantum Information Processing}, vol.~19, pp. 1--8, 2020.

\bibitem{lu2019parameter}
F.-Y. Lu, Z.-Q. Yin, C.~Wang, C.-H. Cui, J.~Teng, S.~Wang, W.~Chen, W.~Huang, B.-J. Xu, G.-C. Guo \emph{et~al.}, ``Parameter optimization and real-time calibration of a measurement-device-independent quantum key distribution network based on a back propagation artificial neural network,'' \emph{JOSA B}, vol.~36, no.~3, pp. B92--B98, 2019.

\bibitem{huang2016long}
D.~Huang, P.~Huang, D.~Lin, and G.~Zeng, ``Long-distance continuous-variable quantum key distribution by controlling excess noise,'' \emph{Scientific reports}, vol.~6, no.~1, p. 19201, 2016.

\bibitem{wang2019machine}
W.~Wang and H.-K. Lo, ``Machine learning for optimal parameter prediction in quantum key distribution,'' \emph{Physical Review A}, vol. 100, no.~6, p. 062334, 2019.

\bibitem{li2018discrete}
J.~Li, Y.~Guo, X.~Wang, C.~Xie, L.~Zhang, and D.~Huang, ``Discrete-modulated continuous-variable quantum key distribution with a machine-learning-based detector,'' \emph{Optical Engineering}, vol.~57, no.~6, pp. 066\,109--066\,109, 2018.

\bibitem{dunjko2018machine}
V.~Dunjko and H.~J. Briegel, ``Machine learning \& artificial intelligence in the quantum domain: a review of recent progress,'' \emph{Reports on Progress in Physics}, vol.~81, no.~7, p. 074001, 2018.

\bibitem{liu2022automated}
Z.-P. Liu, M.-G. Zhou, W.-B. Liu, C.-L. Li, J.~Gu, H.-L. Yin, and Z.-B. Chen, ``Automated machine learning for secure key rate in discrete-modulated continuous-variable quantum key distribution,'' \emph{Optics Express}, vol.~30, no.~9, pp. 15\,024--15\,036, 2022.

\bibitem{mao2020detecting}
Y.~Mao, W.~Huang, H.~Zhong, Y.~Wang, H.~Qin, Y.~Guo, and D.~Huang, ``Detecting quantum attacks: A machine learning based defense strategy for practical continuous-variable quantum key distribution,'' \emph{New Journal of Physics}, vol.~22, no.~8, p. 083073, 2020.

\bibitem{liu2019practical}
J.-Y. Liu, H.-J. Ding, C.-M. Zhang, S.-P. Xie, and Q.~Wang, ``Practical phase-modulation stabilization in quantum key distribution via machine learning,'' \emph{Physical Review Applied}, vol.~12, no.~1, p. 014059, 2019.

\bibitem{liu2018integrating}
W.~Liu, P.~Huang, J.~Peng, J.~Fan, and G.~Zeng, ``Integrating machine learning to achieve an automatic parameter prediction for practical continuous-variable quantum key distribution,'' \emph{Physical Review A}, vol.~97, no.~2, p. 022316, 2018.

\bibitem{chin2021machine}
H.-M. Chin, N.~Jain, D.~Zibar, U.~L. Andersen, and T.~Gehring, ``Machine learning aided carrier recovery in continuous-variable quantum key distribution,'' \emph{npj Quantum Information}, vol.~7, no.~1, p.~20, 2021.

\bibitem{curty2004entanglement}
M.~Curty, M.~Lewenstein, and N.~L{\"u}tkenhaus, ``Entanglement as a precondition for secure quantum key distribution,'' \emph{Physical review letters}, vol.~92, no.~21, p. 217903, 2004.

\bibitem{wang2022numerical}
W.~Wang and N.~L{\"u}tkenhaus, ``Numerical security proof for the decoy-state bb84 protocol and measurement-device-independent quantum key distribution resistant against large basis misalignment,'' \emph{Physical Review Research}, vol.~4, no.~4, p. 043097, 2022.

\bibitem{cai2024free}
W.-Q. Cai, Y.~Li, B.~Li, J.-G. Ren, S.-K. Liao, Y.~Cao, L.~Zhang, M.~Yang, J.-C. Wu, Y.-H. Li \emph{et~al.}, ``Free-space quantum key distribution during daylight and at night,'' \emph{Optica}, vol.~11, no.~5, pp. 647--652, 2024.

\bibitem{scarani2009security}
V.~Scarani, H.~Bechmann-Pasquinucci, N.~J. Cerf, M.~Du{\v{s}}ek, N.~L{\"u}tkenhaus, and M.~Peev, ``The security of practical quantum key distribution,'' \emph{Reviews of modern physics}, vol.~81, no.~3, p. 1301, 2009.

\bibitem{martinez2014demystifying}
J.~Martinez-Mateo, C.~Pacher, M.~Peev, A.~Ciurana, and V.~Martin, ``Demystifying the information reconciliation protocol cascade,'' \emph{arXiv preprint arXiv:1407.3257}, 2014.

\bibitem{sahu2024state}
S.~K. Sahu and K.~Mazumdar, ``State-of-the-art analysis of quantum cryptography: applications and future prospects,'' \emph{Frontiers in Physics}, vol.~12, p. 1456491, 2024.

\bibitem{portmann2022security}
C.~Portmann and R.~Renner, ``Security in quantum cryptography,'' \emph{Reviews of Modern Physics}, vol.~94, no.~2, p. 025008, 2022.

\bibitem{winick2018reliable}
A.~Winick, N.~L{\"u}tkenhaus, and P.~J. Coles, ``Reliable numerical key rates for quantum key distribution,'' \emph{Quantum}, vol.~2, p.~77, 2018.

\bibitem{hu2022robust}
H.~Hu, J.~Im, J.~Lin, N.~L{\"u}tkenhaus, and H.~Wolkowicz, ``Robust interior point method for quantum key distribution rate computation,'' \emph{Quantum}, vol.~6, p. 792, 2022.

\bibitem{mahmud2020variational}
M.~S. Mahmud, J.~Z. Huang, and X.~Fu, ``Variational autoencoder-based dimensionality reduction for high-dimensional small-sample data classification,'' \emph{International Journal of Computational Intelligence and Applications}, vol.~19, no.~01, p. 2050002, 2020.

\bibitem{mertz2013quantum}
M.~Mertz, H.~Kampermann, Z.~Shadman, and D.~Bru{\ss}, ``Quantum key distribution with finite resources: Taking advantage of quantum noise,'' \emph{Physical Review A}, vol.~87, no.~4, p. 042312, 2013.

\bibitem{zhao2019variational}
Q.~Zhao, E.~Adeli, N.~Honnorat, T.~Leng, and K.~M. Pohl, ``Variational autoencoder for regression: Application to brain aging analysis,'' in \emph{Medical Image Computing and Computer Assisted Intervention--MICCAI 2019: 22nd International Conference, Shenzhen, China, October 13--17, 2019, Proceedings, Part II 22}.\hskip 1em plus 0.5em minus 0.4em\relax Springer, 2019, pp. 823--831.

\bibitem{charte2021reducing}
D.~Charte, F.~Charte, and F.~Herrera, ``Reducing data complexity using autoencoders with class-informed loss functions,'' \emph{IEEE Transactions on Pattern Analysis and Machine Intelligence}, vol.~44, no.~12, pp. 9549--9560, 2021.

\bibitem{al2022new}
H.~A. Al-Mohammed and E.~Yaacoub, ``New way to generating and simulation qkd,'' in \emph{Proceedings of Sixth International Congress on Information and Communication Technology}.\hskip 1em plus 0.5em minus 0.4em\relax Springer, 2022, pp. 801--809.

\bibitem{al2021detecting}
H.~A. Al-Mohammed, A.~Al-Ali, E.~Yaacoub, K.~Abualsaud, and T.~Khattab, ``Detecting attackers during quantum key distribution in iot networks using neural networks,'' in \emph{2021 IEEE Globecom Workshops (GC Wkshps)}.\hskip 1em plus 0.5em minus 0.4em\relax IEEE, 2021, pp. 1--6.

\bibitem{calver2011empirical}
T.~Calver, M.~Grimaila, and J.~Humphries, ``An empirical analysis of the cascade error reconciliation protocol for quantum key distribution,'' in \emph{Proceedings of the Seventh Annual Workshop on Cyber Security and Information Intelligence Research}, 2011, pp. 1--1.

\bibitem{timothy2011emprical}
C.~Timothy, ``An emprical analysis of the cascade secret key reconciliation protocol for quantum key distribution,'' Ph.D. dissertation, Master thesis, Air Force Institute of Technology, Air University, 2011.

\bibitem{kingma2014adam}
D.~P. Kingma and J.~Ba, ``Adam: A method for stochastic optimization,'' \emph{arXiv preprint arXiv:1412.6980}, 2014.

\end{thebibliography}

\begin{IEEEbiography}[{\includegraphics[width=1in,height=1.25in,clip,keepaspectratio]{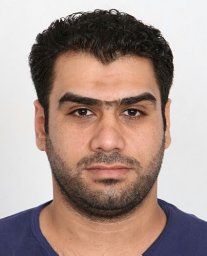}}]{Hasan Abbas Al-Mohammed} received the bachelor’s degree in computer engineering from Iraq University College, Basra, Iraq, in 2014, and the master’s degree in computing from Qatar University, in June 2021. He has more than ten publications in international journals and conferences. His research interests include quantum radar, quantum computing, quantum communications, and security, in addition to the Internet of Things (IoT) and sensor networks. He is currently pursuing his PhD at Qatar University in Quantum Optics.
\end{IEEEbiography}

\begin{IEEEbiography}[{\includegraphics[width=1in,height=1.25in,clip,keepaspectratio]{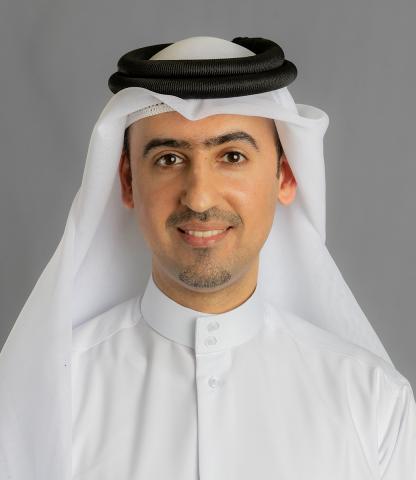}}]{SAIF AL-KUWARI} received a Bachelor of
Engineering in Computers and Networks from
the University of Essex (UK) in 2006 and
two PhD’s from the University of Bath and
Royal Holloway, University of London (UK)
in Computer Science, both in 2012. He
is currently an associate professor at the
College of Science and Engineering at Hamad
Bin Khalifa University. His research interests
include Quantum Computing, Quantum Cryptography and Quantum Machine Learning.
He is IET and BCS fellow, and IEEE and ACM senior member.
\end{IEEEbiography}

\begin{IEEEbiography}
[{\includegraphics[width=1in,height=1.25in,clip,keepaspectratio]{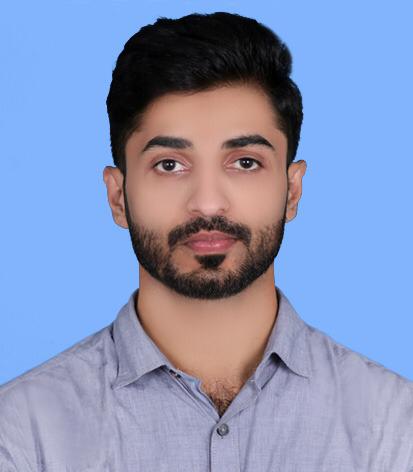}}]
{Hashir Kuniyil} obtained his Master of Science degree in Photonics from Manipal Institute of Technology, Manipal University, India. After which, he worked in the Indian Institute of Technology (IIT) as a Project Associate to carry out a project for the Defense Research and Development Organization (DRDO) of India. Subsequently, he received his PhD in Electrical and Electronics Engineering from Ozyegin University, Istanbul, Turkey, in 2023. Later, he worked as a Quantum Optics Lab Supervisor in the Quantum Optical Lab at the same university. At present, he is a post-doctoral fellow at the College of Science and Engineering, Hamad Bin Khalifa University Doha, Qatar. His research interests include photonic quantum imaging, experimental quantum key distribution, and atomic-physics experiments for quantum sensing.
\end{IEEEbiography}
\begin{IEEEbiography}[{\includegraphics[width=1in,height=1.25in,clip,keepaspectratio]{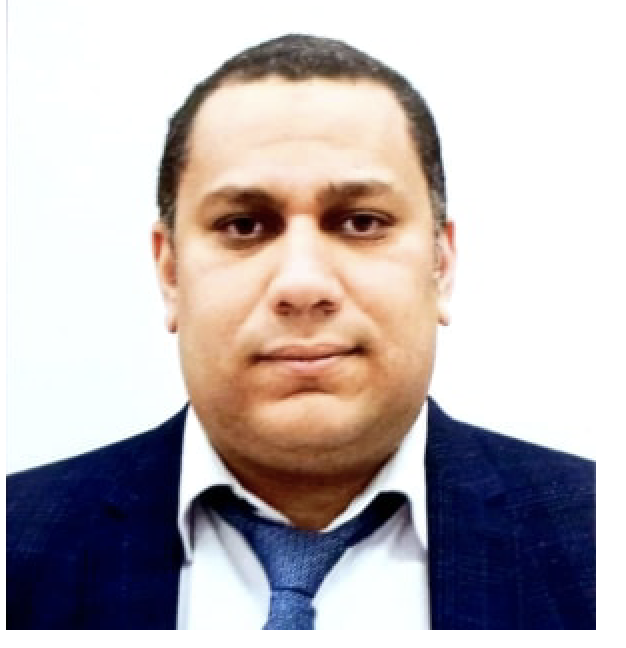}}]{Ahmed Farouk} is currently a senior scientist at Qatar Center of Quantum Computing, HBKU, Qatar and an assistant professor at the faculty of computers and artificial intelligence at South Valley University, Egypt, and an early career scientist demonstrating excellence in quantum computing and cybersecurity research by publishing more than 150 research papers with a high impact. Most are published in top venues such as IEEE WCM, IEEE TAI, IEEE IOTJ, IEEE TITS, IEEE TII, IEEE TNSE, and IEEE TIV. He is becoming internationally recognized as a leading researcher. This is clear from receiving the Egyptian Encouragement Award in advanced technological sciences, his selection as one of 17 researchers from Africa to attend the 70th Lindau Meeting of Nobel Laureates, won travel grants by IEEE Computer Society and Okinawa Institute of Science and Technology. His volunteering effort is apparent in working as a reviewer for more than 40 journals, associate editor member for many journals, Track and Session Chair for conferences. Furthermore, he chaired the IEEE Computer Society Chapter and was elected as an officer for the Consumer Technology Society (CTSoc) on Quantum Consumer Technology Technical Committee (QCT).
%\end{IEEEbiography
\end{IEEEbiography}
 
\end{document}